\documentclass[reqno,11pt]{amsart}
\usepackage{amsmath,amssymb,amsthm,graphicx,a4wide,enumerate,url}
\usepackage[small,bf]{caption} 
\newcommand{\prn}[1]{\left(#1\right)}
\newcommand{\abs}[1]{\left|#1\right|}

\newcommand{\ud}[1]{\,\mathrm{d}#1}

\begin{document}
\parskip.9ex

\title[Effect of the Choice of Stagnation Density in Data-Fitted Traffic Models]
{Effect of the Choice of Stagnation Density in Data-Fitted First- and Second-Order Traffic Models}
\author[S. Fan]{Shimao Fan}
\address[Shimao Fan]
{Department of Mathematics \\ Temple University \\ \newline
1805 North Broad Street \\ Philadelphia, PA 19122}
\email{shimao.fan@temple.edu}
\author[B. Seibold]{Benjamin Seibold}
\address[Benjamin Seibold]
{Department of Mathematics \\ Temple University \\ \newline
1805 North Broad Street \\ Philadelphia, PA 19122}
\email{seibold@temple.edu}
\urladdr{http://www.math.temple.edu/\~{}seibold}

\begin{abstract}
For a class of data-fitted macroscopic traffic models, the influence of the choice of the stagnation density on the model accuracy is investigated. This work builds on an established framework of data-fitted first-order Lighthill-Whitham-Richards (LWR) models and their second-order Aw-Rascle-Zhang (ARZ) generalizations. These models are systematically fitted to historic fundamental diagram data, and then their predictive accuracy is quantified via a version of the three-detector problem test, considering vehicle trajectory data and single-loop sensor data. The key outcome of this study is that with commonly suggested stagnation densities of 120 vehicles/km/lane and above, information travels backwards too slowly. It is then demonstrated that the reduction of the stagnation density to 90--100 vehicles/km/lane addresses this problem and results in a significant improvement of the predictive accuracy of the considered models.
\end{abstract}

\maketitle

\section{Introduction}
\label{sec:introduction}
This paper builds on a framework to construct data-fitted macroscopic models and to test their predictive accuracy, presented in \cite{FanSeibold2013}. In said paper, a flow-density function $Q(\rho)$ is least-squares fitted to historic fundamental diagram data, where the solution is drawn from a three-parameter family of functions, all of which vanish for $\rho = 0$ and for the \emph{stagnation density} $\rho = \rho_\text{max}$. Because the latter is commonly badly represented in measurement data (see the data points in Figs.~\ref{fig:evolution_ngsim} and~\ref{fig:evolution_rtmc}), it is prescribed as a fixed constant, independent of the particular data. This is justified because the safety distance that vehicles keep when coming to a complete stop is independent of the dynamics of moving traffic flow. In \cite{Helbing2001} it is suggested that in traffic engineering one typically chooses $\rho_\text{max}\in [120,200]\; \text{veh}/\text{km}/\text{lane}$. In the study conducted in \cite{FanSeibold2013}, the choice is $\rho_\text{max} = 133.33\,\text{veh}/\text{km}/\text{lane}$, which corresponds to a vehicle length of 5m plus 50\% safety distance.

In this paper, we investigate in which way the choice of $\rho_\text{max}$ influences the predictive accuracy of the data-fitted models, i.e., how well they reproduce the true behavior of traffic flow. This was initially motivated as a simple study of the robustness of the results in \cite{FanSeibold2013} with respect to the choice of this model parameter. However, the results have turned out to highlight at a much deeper result, namely that the traffic models with $\rho_\text{max}\in [120,200]\; \text{veh}/\text{km}/\text{lane}$ have information propagating too slowly backwards on the road; and that choosing $\rho_\text{max}\in [90,100]\; \text{veh}/\text{km}/\text{lane}$ leads to much more realistic information propagation speeds and consequently to more accurate model predictions.

The methodology to conduct the data-fitting and the model validation is very similar to the approaches in \cite{FanSeibold2013}. Two different types of data are considered: vehicle trajectory data (NGSIM data set \cite{TrafficNGSIM_I80}), and single-loop sensor data (RTMC data set \cite{TrafficMnDOT}). Each type of data comes with historic fundamental diagram data that is used to determine two first-order models that differ in the type of flux function used, see \S\ref{sec:model_creation}. As described in \S\ref{sec:traffic_models}, these models generalize to second-order models, which possess a family of flow-density functions. The predictive accuracy of the models is then determined via a variant of the three-detector problem \cite{Daganzo1997}: on the considered segment of highway, the traffic state is made available to the model at both ends (at all times), and the states that the model creates inside the study domain are then compared with the real data at the same positions and times. The model validation is conducted in a fully macroscopic sense, i.e., \vspace{-.6em}
\begin{enumerate}[a)]
\item the discrete data (vehicle trajectories or aggregated sensor data) are transformed into field quantities that are defined continuously in space and time, see \S\ref{sec:description_data};
\item the numerical schemes used to approximate the governing PDE are applied on very fine computational grids, so that the numerical approximation errors are negligibly small relative to the model errors (see \S\ref{subsec:model_computation}); and
\item differences between model predictions and measurement data are evaluated in an $L^1$ fashion, i.e., integrated over space and/or time (see \S\ref{subsec:error_measures}).
\end{enumerate}

Relative to the preceding work \cite{FanSeibold2013}, two technical improvements are contributed: one in the preprocessing of the NGSIM data (see \S\ref{subsec:data_ngsim}), and another in the error measure used to quantity the model accuracy (see \S\ref{subsec:error_measures}). The central results are then presented in \S\ref{sec:results_wave_speed} in the form of a phenomenological study of how well different models reproduce real wave propagation speeds, and in \S\ref{sec:results_model_accuracy} via the investigation of how the predictive accuracy of various models depends on the choice of $\rho_\text{max}$.

\vspace{1.5em}
\section{Macroscopic Traffic Models}
\label{sec:traffic_models}
Macroscopic models use partial differential equations to describe the temporal evolution of the (lane-aggregated) traffic density $\rho(x,t)$ that is defined at every position $x$ along the road. There is a wide variety of alternative ways to describe vehicular traffic flow, most prominently microscopic models (e.g., \cite{Pipes1953, Newell1961}), mesoscopic models (e.g., \cite{HermanPrigogine1971, Phillips1979}), probabilistic and cellular models (e.g., \cite{NagelSchreckenberg1992}). These various types of descriptions are related. For instance, macroscopic models become cell transmission models when discretized in Eulerian variables \cite{Daganzo1994}, and they become microscopic models when discretized in Lagrangian variables \cite{AwKlarMaterneRascle2002}. Through these relations, the results obtained here also have some relevance for other types of models.

In traffic engineering applications, macroscopic models have been successfully used to incorporate measurement data into a running computation, and have particular strengths when those data have a low penetration (e.g., \cite{WangPapageorgiou2005, Aminetal2008, HerreraWorkBanHerringJacobsonBayen2010, BlandinCoqueBayen2012}). In these applications, the models are commonly combined with a filtering framework (see \cite{Kalman1960, MihaylovaBoelHegyi2007}), i.e., the model parameters are modified over the course of the computation. In contrast, in this study all model parameters are determined a-priori (from historic fundamental diagram data), and then the model is validated with time-dependent data. The reason is that the goal here is to study the ability of various models to reproduce the behavior of real traffic, and thus to determine which types of models would be suitable candidates for practical applications (i.e., in combination with filtering).

\subsection{First- and Second-Order Macroscopic Models}
\label{subsec:first_and_second_order_models}
Common to all macroscopic models is the continuity equation
\begin{equation}
\label{eq:continuity_equation}
\rho_t+(\rho u)_x = 0\;,
\end{equation}
that encodes the conservation of vehicles that move under the velocity field $u(x,t)$. The Lighthill-Whitham-Richards (LWR) model \cite{LighthillWhitham1955, Richards1956} provides a closure by prescribing a functional relationship between $\rho$ and $u$ via a velocity function, $u = U(\rho)$. This induces a flow rate function $Q(\rho) = \rho U(\rho)$, that acts as the flux function in the LWR model
\begin{equation}
\label{eq:lighthill_whitham_richards_model}
\rho_t+(Q(\rho))_x = 0\;.
\end{equation}
which is a scaler hyperbolic conservation law and thus called a \emph{first-order model}.

\emph{Second-order models} add an evolution equation for the velocity field $u(x,t)$ to \eqref{eq:continuity_equation}, thus leading to $2\times 2$ systems of hyperbolic conservation laws. A variety of second-order models has been proposed, most prominently the Payne-Whitham (PW) model \cite{Payne1971, Whitham1974} and the Aw-Rascle-Zhang (ARZ) model \cite{AwRascle2000, Zhang2002}. Here, we consider the latter; specifically, the homogeneous ARZ model, which reads as
\begin{equation}
\label{eq:aw_rascle_zhang_model_homogeneous}
\begin{split}
\rho_t+(\rho u)_x &= 0\;, \\
(u+h(\rho))_t+u(u+h(\rho))_x &= 0\;.
\end{split}
\end{equation}
It should be remarked that \eqref{eq:aw_rascle_zhang_model_homogeneous}, and its reformulations below, are written in primitive variables $(\rho,u)$ for convenience. However, their weak solutions are defined via the conservative form which follows by rewriting \eqref{eq:aw_rascle_zhang_model_homogeneous} in terms of the variables $(\rho,\rho (u+h(\rho))$, see \cite{AwRascle2000, FanSeibold2013}. Further note that one could also consider an inhomogeneous ARZ model \cite{AwRascle2000}, which possesses a relaxation term in the velocity equation. This generalization shall be presented in detail in an upcoming paper \cite{FanHertySeibold2013}.

The function $h(\rho)$ in \eqref{eq:aw_rascle_zhang_model_homogeneous} is called the \emph{hesitation function} (sometimes also called ``pressure''), and we assume that $h'(\rho)>0$ and $h(0) = 0$. The reason for this terminology is that system \eqref{eq:aw_rascle_zhang_model_homogeneous} can be rewritten as
\begin{equation}
\label{eq:aw_rascle_zhang_model_w}
\begin{split}
\rho_t+(\rho u)_x &= 0\;, \\
w_t+uw_x &= 0\;, \\
\text{where~}u &= w-h(\rho)\;.
\end{split}
\end{equation}
Because the quantity $w = u+h(\rho)$ is advected with the velocity field $u$, it can be interpreted as a property associated with each vehicle. Specifically, because $w = u$ for $\rho = 0$, we interpret $w$ as the \emph{empty road velocity} that a driver would assume when alone on the road. The actual velocity $u$ is then given by $w$, reduced by the hesitation $h(\rho)$. With this interpretation the ARZ model is a natural generalization of the LWR model \eqref{eq:lighthill_whitham_richards_model}, namely: instead of a single velocity vs.~density curve $U(\rho)$, it possesses a one-parameter family of velocity vs.~density curves $u_w(\rho) = w-h(\rho)$. With the choice $h(\rho) = U(0)-U(\rho)$, the ARZ model is defined via a family of velocity curves
\begin{equation}
\label{eq:arz_generalized_u}
u_w(\rho) = U(\rho)+(w-U(0))\hspace{.85em}
\end{equation}
and resulting density curves
\begin{equation}
\label{eq:arz_generalized_Q}
Q_w(\rho) = Q(\rho)+\rho(w-U(0))\;.
\end{equation}
In other words, the ARZ model is based on the same model parameter function $U(\rho)$ as the LWR model; but it generalizes it to allow drivers to have different velocities while being at the same density $\rho$ (thus modeling different types of drivers). In the $\rho$--$u$ diagram, the curves \eqref{eq:arz_generalized_u} are just vertical shifts of $U(\rho)$; and in the $\rho$--$Q$ diagram, different linear functions are added to $Q(\rho)$. This is shown in the left panels of Figs.~\ref{fig:evolution_ngsim} and~\ref{fig:evolution_rtmc}, in which the black curves represent the ARZ family $Q_w(\rho)$, and the red curve is the specific LWR function $Q(\rho)$.

Since the ARZ model \eqref{eq:aw_rascle_zhang_model_w} does not possess more modeling parameters than the LWR model \eqref{eq:lighthill_whitham_richards_model}, one may wonder whether it actually reproduces real traffic behavior better. The results in \cite{FanSeibold2013} and in this paper confirm: yes, it does. The main reasons are as follows. First, LWR's stringent coupling between density and velocity is relaxed by the ARZ model, and the set-valued nature of real fundamental diagrams (cf.~\cite{Kerner1998, Helbing2001}) can be captured via the family of curves \eqref{eq:arz_generalized_Q}. Second, in first-order models, drivers are assumed to adjust instantaneously to changes in density, while second-order models possess an actual model for acceleration and deceleration. And third, second-order models allow for the incorporation of more data through the boundaries of the computational domain.

\subsection{Wave Propagation Speed in the Traffic Models}
\label{subsec:wave_speed}
In macroscopic models, information can propagate by two means: first, along characteristics where the solution is continuous; and second, via shocks and contacts where the solution is discontinuous. Both the LWR and the ARZ models have a characteristic velocity that is slower than the vehicles; in addition, the ARZ model possesses a second characteristic that moves with the vehicles and that is associated with contact discontinuities. Moreover, both models possess genuinely nonlinear waves, i.e., shocks and contact discontinuities, that move slower than the vehicles. The speed of shocks is given by the Rankine-Hugoniot jump conditions.

In both types of traffic models, the travel speed of information can be extracted graphically from the fundamental diagram curves, as given in the left panels of Figs.~\ref{fig:evolution_ngsim} and~\ref{fig:evolution_rtmc}. For the LWR model \eqref{eq:lighthill_whitham_richards_model} the characteristic velocity, $\lambda = Q'(\rho)$, is given by the tangent slope to the red curve $Q(\rho)$; and the speed of shocks, $s = [Q(\rho)]/[\rho]$, is given by the slope of a secant to the red curve. Here, brackets denote the jump of a quantity across the shock. Analogously, for the ARZ model \eqref{eq:aw_rascle_zhang_model_homogeneous} the slower characteristic velocity, $\lambda_1 = Q_w'(\rho)$, is given by the tangent slope to any of the black curves $Q_w(\rho)$; and the speed of shocks, $s = [Q_w(\rho)]/[\rho]$, is given by the secant slope to any of the black curves (see \cite{AwRascle2000} for the derivation of these formulae). Therefore, the steeper the decrease of the fundamental diagram curves, the faster information travels backwards in the models. This fundamental property is important for the interpretation of the results found in \S\ref{sec:results_wave_speed}.

\vspace{1.5em}
\section{Data Sets Used and Their Preprocessing}
\label{sec:description_data}
As in \cite{FanSeibold2013}, the study is conducted using two different types of data: trajectory data, as described in \S\ref{subsec:data_ngsim}, and single-loop sensor data, as described in \S\ref{subsec:data_rtmc}.

\subsection{Processing of the NGSIM Trajectory Data}
\label{subsec:data_ngsim}
The NGSIM vehicle trajectory data set \cite{TrafficNGSIM_I80} was collected in 2005 on a $\approx\! 500\text{m}$ long segment of the eastbound direction of I-80 located in Emeryville, CA. In three 15-minute intervals, the trajectories of all vehicles in the segment are available (with a temporal resolution of 0.1 seconds). In addition, traditional fundamental diagram data, obtained via loop detectors, is available for the same highway segment \cite{TrafficNGSIM_I80}.

For the use in our macroscopic study, the NGSIM data are processed as described in \cite{FanSeibold2013}, with a few minor modifications. First, all trajectories are projected onto a single lane, neglecting vehicles on the on-ramp that is in the segment. Second, vehicle trajectories that enter after $22\text{m}$, and not from the ramp, are extended backwards in time. Third, all data outside the interval $[22\text{m},500\text{m}]$ are removed. Fourth, at any instance in time, a kernel density estimation (KDE) is applied to the vehicle positions, as described below. Fifth, the resulting density and velocity fields are restricted to the domain $[36\text{m},486\text{m}]$. Sixth, the data are smoothed in time via a least squares approach, see below.

The KDE \cite{Rosenblatt1956} on the data is conducted as in \cite{FanSeibold2013}. Let at any time step the vehicle positions $x_1,\dots,x_N$ and their velocities $u_1,\dots,u_N$ be given. First, these data are extended via the reflection method described in \cite{FanSeibold2013}, i.e., suitable ghost vehicles are added to the left of $x_1$ and to the right of $x_N$, leading to the extended vehicle positions $\hat{x}_1,\dots,\hat{x}_{\hat{N}}$ and velocities $\hat{u}_1,\dots,\hat{u}_{\hat{N}}$. Second, the vehicle density field and the associated flow rate field are reconstructed as superpositions of Gaussians
\begin{equation}
\label{eq:kde}
\rho(x) = \sum_{j=1}^N K(x-\hat{x}_j)
\quad\text{and}\quad
Q(x) = \sum_{j=1}^N u_j K(x-\hat{x}_j)\;,
\text{~with}\quad
K(x) = \frac{1}{\sqrt{2\pi}h}e^{-\frac{x^2}{2h^2}}\;,
\end{equation}
where the kernel width is chosen as $h = 25\text{m}$. Using the two fields in \eqref{eq:kde}, the velocity field is then defined as $u(x) = Q(x)/\rho(x)$.

Since the KDE reconstruction is applied at any instance in time (in 0.1\text{s} intervals), evolving density and velocity fields, $\rho(x,t)$ and $u(x,t)$ are defined. The temporal intervals for the study must be chosen so that at any instance in time, the complete study interval $x\in [36\text{m},486\text{m}]$ is filled with tracked vehicles. The resulting time intervals are: $t\in [\text{4:00:30pm},\text{4:14:00pm}]$ for the first data set, $t\in [\text{5:00:30pm},\text{5:13:30pm}]$ for the second, and $t\in [\text{5:15:30pm},\text{5:28:00pm}]$ for the third data set.

Finally, a temporal smoothing procedure is applied to remedy spurious fast-scale oscillations (in time) near the boundaries (due to the discontinuous behavior of the data when vehicles pop in or pop out of the data set). The time evolution of a field quantity, say $\rho(x,\cdot)$ at a given position $x$, is replaced by an approximating cubic spline that is globally $C^1$. The time interval is divided into segments of length $\Delta t_\text{s} = 15\text{s}$. On each segment $[t_n,t_n+\Delta t_\text{s}]$, the cubic polynomial is defined as the least-squares approximation to the 150 data points on that interval, with the constraints that the value and slope at $t_n$ are matched to those obtained in the previous interval (in $[0,\Delta t_\text{s}]$, an unconstrained LSQ-fit is conducted). The segment length $\Delta t_\text{s}$ is chosen, so that the results in the interior of the study domain are minimally affected by the smoothing, while the spurious oscillations near the boundaries are removed.

\subsection{Processing of the RTMC Sensor Data}
\label{subsec:data_rtmc}
The second data set is a part of the RTMC data \cite{TrafficMnDOT} from the year 2003, provided by Mn/DOT. The data is obtained via single loop sensors, which measure traffic volume an occupancy at fixed positions on the highway, aggregated over intervals of $\Delta t_\text{a} = 30\text{s}$. As described in \cite{FanSeibold2013}, vehicle densities and velocities are constructed from the measured quantities. Three successive sensor positions (denoted sensors 1, 2, and 3) are considered, along the southbound direction of I-35W, south of its intersection with I-94. The study segment (between sensors 1 and 3) is of length 1,224m. For this study, the hour 4pm--5pm on 74 weeks days (Monday--Friday) within 01/01--04/16/2003 is considered, and out of these, 43 ``congested days'' are selected, defined as the average traffic density between 4pm and 5pm exceeding $20\,\text{veh}/\text{km}/\text{lane}$.

At the three sensor positions, continuous-in-time density and velocity functions are defined as follows. On each aggregation interval $[t_n,t_n+\Delta t_\text{a}]$, the respective quantities are assigned to the mid-time $t_n+\frac{1}{2}\Delta t_\text{a}$, and then a cubic spline interpolant is defined on $t\in [4\text{pm},5\text{pm}]$ based on these data points. In addition to these time-dependent data, a fundamental diagram is generated from the data at sensor 2 (aggregated over intervals of 10 minutes), recorded over the whole year 2003.

\vspace{1.5em}
\section{Model Creation via Fitting to Fundamental Diagram Data}
\label{sec:model_creation}
As in \cite{FanSeibold2013}, we consider a class of flow rate functions that are smooth and strictly concave down ($Q''(\rho)<0$) everywhere, and that vanish for $\rho = 0$ and $\rho = \rho_\text{max}$, given by the expression
\begin{equation}
\label{eq:flow_rate_vs_density_function}
Q(\rho) = \alpha\prn{a+(b-a)\tfrac{\rho}{\rho_{\text{max}}}-\sqrt{1+y^2}}\;,
\end{equation}
where
\begin{equation}\nonumber
a = \sqrt{1+(\lambda p)^2}\;,\quad
b = \sqrt{1+(\lambda(1-p))^2}\;,\text{~and}\quad
y = \lambda\prn{\tfrac{\rho}{\rho_{\text{max}}}-p}\;.
\end{equation}
The three parameters $\alpha$, $\lambda$, and $p$ control the critical density $\rho_\text{c}$ (at which $Q'(\rho_\text{c}) = 0$), the maximum flow rate $Q(\rho_\text{c})$, and the ``roundness'' of $Q$, i.e., how rapidly the slope $Q'(\rho)$ transitions from positive to negative near $\rho_\text{c}$. This function \eqref{eq:flow_rate_vs_density_function} can be interpreted as a smooth and strictly concave variant of the piecewise-linear Daganzo-Newell (DN) flux \cite{Newell1993, Daganzo1994}. The main reason for not using the simpler DN flux here is that it would render the ARZ model \eqref{eq:aw_rascle_zhang_model_homogeneous} not hyperbolic near $\rho = 0$ (see \cite{FanHertySeibold2013}).

Given fundamental diagram (FD) data, i.e., pairs $(\rho_j,Q_j)$ for $j=1,\dots,n$, the model parameters in \eqref{eq:flow_rate_vs_density_function} are determined as follows. First the stagnation density $\rho_{\text{max}}$ is fixed (in \cite{FanSeibold2013} it was uniformly chosen $\rho_\text{max} = 133.33\,\text{veh}/\text{km}/\text{lane}$; here we vary it over a range of $\rho_\text{max}\in [60,200]\; \text{veh}/\text{km}/\text{lane}$). Then, the three parameters $\alpha$, $\lambda$, and $p$ are determined so that the flow rate function \eqref{eq:flow_rate_vs_density_function} best approximates the FD data in a least-squares sense, i.e., we solve the minimization problem
\begin{equation}\nonumber
\min_{\alpha,\lambda,p}\,\sum_{j=1}^n\abs{Q_{\alpha,\lambda,p}(\rho_j)-Q_j}^2\;.
\end{equation}

In addition to the three-parameter function \eqref{eq:flow_rate_vs_density_function}, we also investigate models based on the Greenshields flux \cite{Greenshields1935}
\begin{equation}
\label{eq:flow_rate_vs_density_function_Greenshields}
Q(\rho) = \rho\,u_\text{max}(1-\rho/\rho_\text{max})\;,
\end{equation}
whose quadratic form was suggested based on Greenshields's first FD measurements in the 1930s. Even though it is now well-known that the Greenshields flux is not a good representation of real FD data, due to it simplicity it is still a popular choice in traffic engineering. It is for this reason that we include it in our investigation here.

Because the Greenshields flux \eqref{eq:flow_rate_vs_density_function_Greenshields} does not resemble the basic shape of FD data well, it is not recommendable to obtain its model parameters via a LSQ-fit. Instead, given $\rho_{\text{max}}$, we determine its empty road velocity $u_\text{max}$ as follows. Given the LSQ-fitted three-parameter function \eqref{eq:flow_rate_vs_density_function}, we choose $u_\text{max} = Q'_{\alpha,\lambda,p}(0)$, so that the slopes of the two flow rate curves match at the origin.

Following the derivation in \S\ref{sec:traffic_models}, the two types of flow rate functions \eqref{eq:flow_rate_vs_density_function} and \eqref{eq:flow_rate_vs_density_function_Greenshields} define two first-order LWR models \eqref{eq:lighthill_whitham_richards_model}, and also two second-order ARZ models \eqref{eq:aw_rascle_zhang_model_w}, whose families of flow rate curves \eqref{eq:arz_generalized_Q} are inherited from the LWR flow rate curves. Hence, for any choice of stagnation density $\rho_\text{max}$, we have four models: ``LWR'' and ``ARZ'', based on \eqref{eq:flow_rate_vs_density_function}, and ``LWRQ'' and ``ARZQ'', based on the quadratic function \eqref{eq:flow_rate_vs_density_function_Greenshields}.

\vspace{1.5em}
\section{Model Validation Procedure and Model Error Quantification}
\label{sec:model_validation}

\subsection{Computation of the Model Predictions}
\label{subsec:model_computation}
Both the LWR model \eqref{eq:lighthill_whitham_richards_model} and the ARZ model \eqref{eq:aw_rascle_zhang_model_homogeneous} are (systems of) hyperbolic conservation laws of the form
\begin{equation}
\label{eq:conservation_law}
\phi_t+f(\phi)_x = 0\;,
\end{equation}
where $\phi = \rho$ and $f = Q$ for LWR; and $\phi = (\rho,q)$ and $f(\rho,q) = (q-\rho h(\rho),q^2/\rho-h(\rho)q)$ for ARZ. The solution $\phi(x,t)$ is defined on a computational domain $(x,t)\in [x_\text{L},x_\text{R}]\times [0,t_\text{f}]$, where $t_\text{f}$ is the final time of the computation. Hence, an initial condition $\phi(x,0)$ is required, as well as boundary conditions at $x_\text{L}$ and $x_\text{R}$ for basic waves that enter the domain.

Due to the construction in \S\ref{subsec:data_ngsim}, for the NGSIM data, initial and boundary conditions conditions are available. For the RTMC data, boundary conditions are available (see \S\ref{subsec:data_rtmc}); however, initial conditions are not (because the traffic state is only known at the sensor positions). This lack of information is overcome as follows. Before the model comparison is conducted, the traffic model is run through an initialization phase (about 5 minutes long) which is started with a uniform state. During this phase, correct boundary conditions are prescribed, and thus the traffic model ``fills'' the interval with a realistic state everywhere in the domain.

The PDE \eqref{eq:conservation_law} is approximated via a standard Godunov method (see \cite{LeVeque2002} for an overview of suitable numerical methods). In order to ensure that the model accuracy of the macroscopic description is measured, a very fine spatial grid resolution is chosen ($\Delta x\le 50\text{cm}$), so that the numerical approximation errors are much smaller than the model errors. The time step size $\Delta t$ is selected so that the CFL condition, $s_{\text{max}}\Delta t \leq \Delta x$, is satisfied, where $s_{\text{max}}$ is the largest wave speed. This fine grid resolution also ensures that any spurious overshoots that may occur in the velocity (cf.~\cite{ChalonsGoatin2007}) are small.

In the Godunov scheme, boundary conditions are provided as follows. On each side of the domain, i.e., at $[x_\text{L}-\Delta x,x_\text{L}]$ and at $[x_\text{R},x_\text{R}+\Delta x]$, a ghost cell is added, in which the boundary state is specified. Thus, the Riemann solver that is invoked at the boundaries $x_\text{L}$ and $x_\text{R}$ selects precisely the right amount and type of boundary information that the solution calls for (see \cite{LeVeque2002} for more details). Specifically, the LWR model \eqref{eq:lighthill_whitham_richards_model} uses density information only, while the ARZ model \eqref{eq:aw_rascle_zhang_model_homogeneous} uses density and velocity information. For light traffic, LWR has $\rho$ enter the domain at $x_\text{L}$ and ARZ has $\rho$ and $u$ enter, while no information enters at $x_\text{R}$. In contrast for dense traffic, LWR has $\rho$ enter at $x_\text{R}$ only, while ARZ has some combination of $\rho$ and $u$ enter the domain at both boundaries.

One further aspect must be considered. The flow rate functions for LWR \eqref{eq:flow_rate_vs_density_function} and LWRQ \eqref{eq:flow_rate_vs_density_function_Greenshields} are only defined for $\rho\le\rho_\text{max}$. Therefore, whenever in the data $\rho$ exceeds $\rho_\text{max}$ (which occurs when $\rho_\text{max}$ is small) it is capped at $\rho_\text{max}$.

\subsection{Error Measures of the Model Accuracy}
\label{subsec:error_measures}
All traffic models produce predictions of the traffic state $(\rho,u)$ in the whole space-time domain $(x,t)\in [x_\text{L},x_\text{R}]\times [0,t_\text{f}]$, where first-order models yield estimates for the velocity via the unique velocity vs.~density relation $u = U(\rho)$.

Our goal is to obtain models that yield good predictions for traffic densities and velocities. Therefore, we define an error measure that involves both density errors, $|\rho^\text{data}-\rho^\text{model}|$, and velocity errors, $|u^\text{data}-u^\text{model}|$. Since these two expressions have different physical units, they must be scaled by a reference density and a reference velocity. In the preceding study \cite{FanSeibold2013}, density errors were divided by $\rho_\text{max}$ and velocity errors by $u_\text{max} = U(0)$. However since, even for dense but moving traffic, velocities tend to be not much less than $\frac{1}{2}u_\text{max}$, while densities tend to be much smaller than $\frac{1}{2}\rho_\text{max}$, one can argue that velocity errors were given a larger weight than density errors.

We therefore propose a new error measure that we argue gives a fair weight to errors in density and in velocity, namely
\begin{equation}
\label{eq:error_e}
E(x,t) = \frac{\abs{\rho^\text{data}(x,t)-\rho^\text{model}(x,t)}}{\Delta\rho}
+ \frac{\abs{u^\text{data}(x,t)-u^\text{model}(x,t)}}{\Delta u}\;.
\end{equation}
Here, $\Delta\rho$ is the maximum variation in density that the historic FD data exhibits, and $\Delta u$ is the maximum variation in velocity, both modulo outliers. To define these two quantities, we consider the FD data $(\rho_j,Q_j)$ for $j=1,\dots,n$, and conduct the following steps. First, we neglect all data with $\rho_j < 5\,\text{veh}/\text{km}/\text{lane}$, i.e., the big chunk of data that represents night time hours, at which a macroscopic description is not justified. Second, $\Delta\rho$ is defined such that 99.9\% of the remaining data points have a lower density. Third, $u^\text{high}$ and $u^\text{low}$ are defined such that 99.9\% of the remaining data points have a lower (or higher, respectively) velocity. Third, we set $\Delta u = u^\text{high}-u^\text{low}$.

The error \eqref{eq:error_e} is defined wherever data is available. For the NGSIM data set this is everywhere in the space-time domain. In contrast, the RTMC data is only defined at the position of the middle sensor, $x_2$. However, the RTMC data covers 43 days for which a significant congestion level is reached during rush hour. We therefore define total model errors as follows. For each of the NGSIM data sets we average over space and time
\begin{equation}
\label{eq:error_xt}
E = \frac{1}{x_\text{R}-x_\text{L}}\int_{x_\text{L}}^{x_\text{R}}
\frac{1}{t_\text{f}}\int_0^{t_\text{f}} E(x,t) \ud{t} \ud{x}\;,
\end{equation}
and for the RTMC data we average over time and over the collection of days in the study
\begin{equation}
\label{eq:error_days_t}
E = \frac{1}{\text{\#days}}\sum_{j=1}^{\text{\#days}}
\frac{1}{t_\text{f}}\int_0^{t_\text{f}} E_\text{day $j$}(x_2,t) \ud{t}\;.
\end{equation}
Since the model predictions are computed on very fine grid resolutions, we can approximate these integrals with negligible errors.

\begin{figure}[p]
\begin{minipage}[b]{.412\textwidth}
\includegraphics[width=\textwidth]{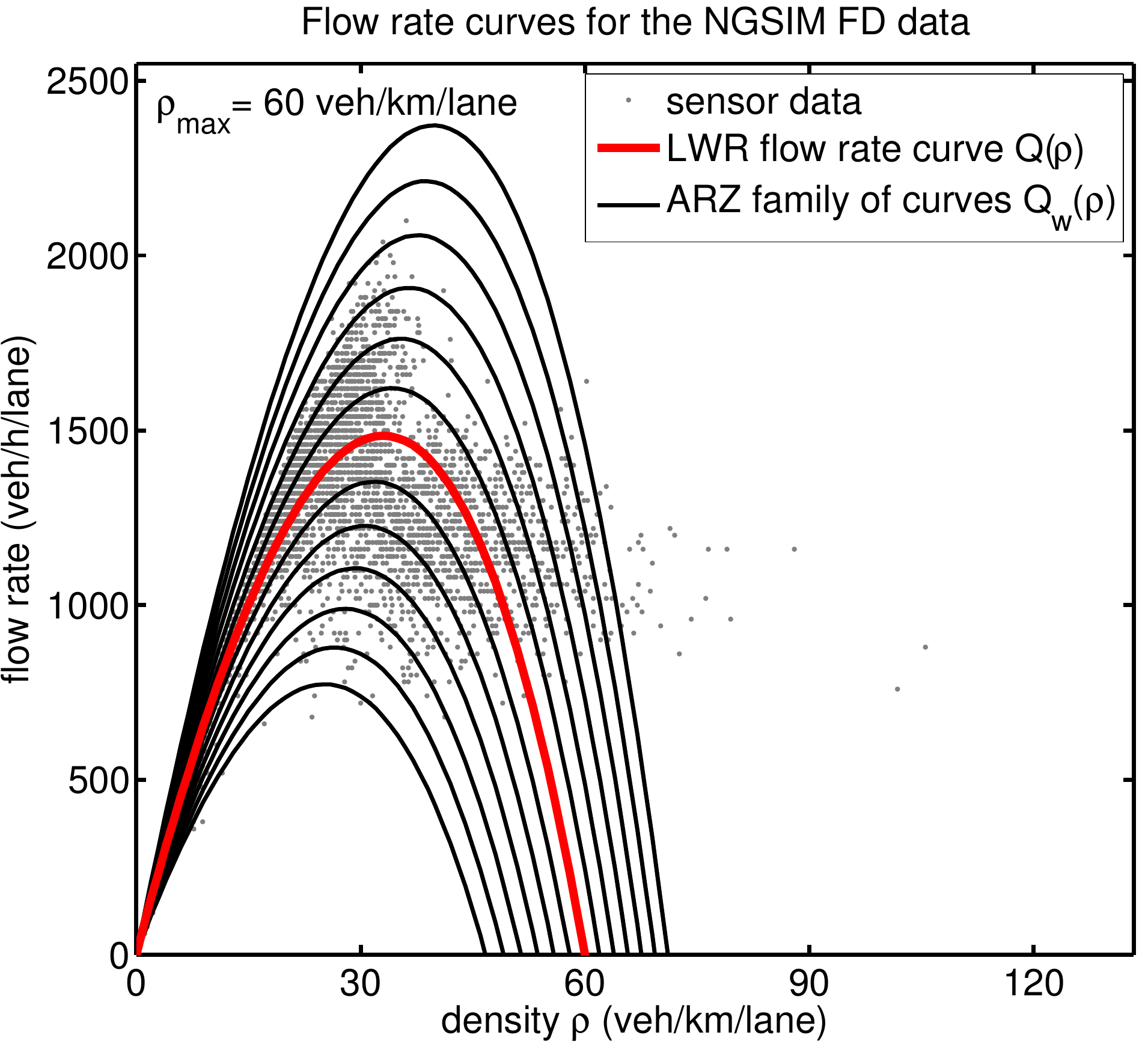}
\end{minipage}
\hfill
\begin{minipage}[b]{.56\textwidth}
\includegraphics[width=\textwidth]{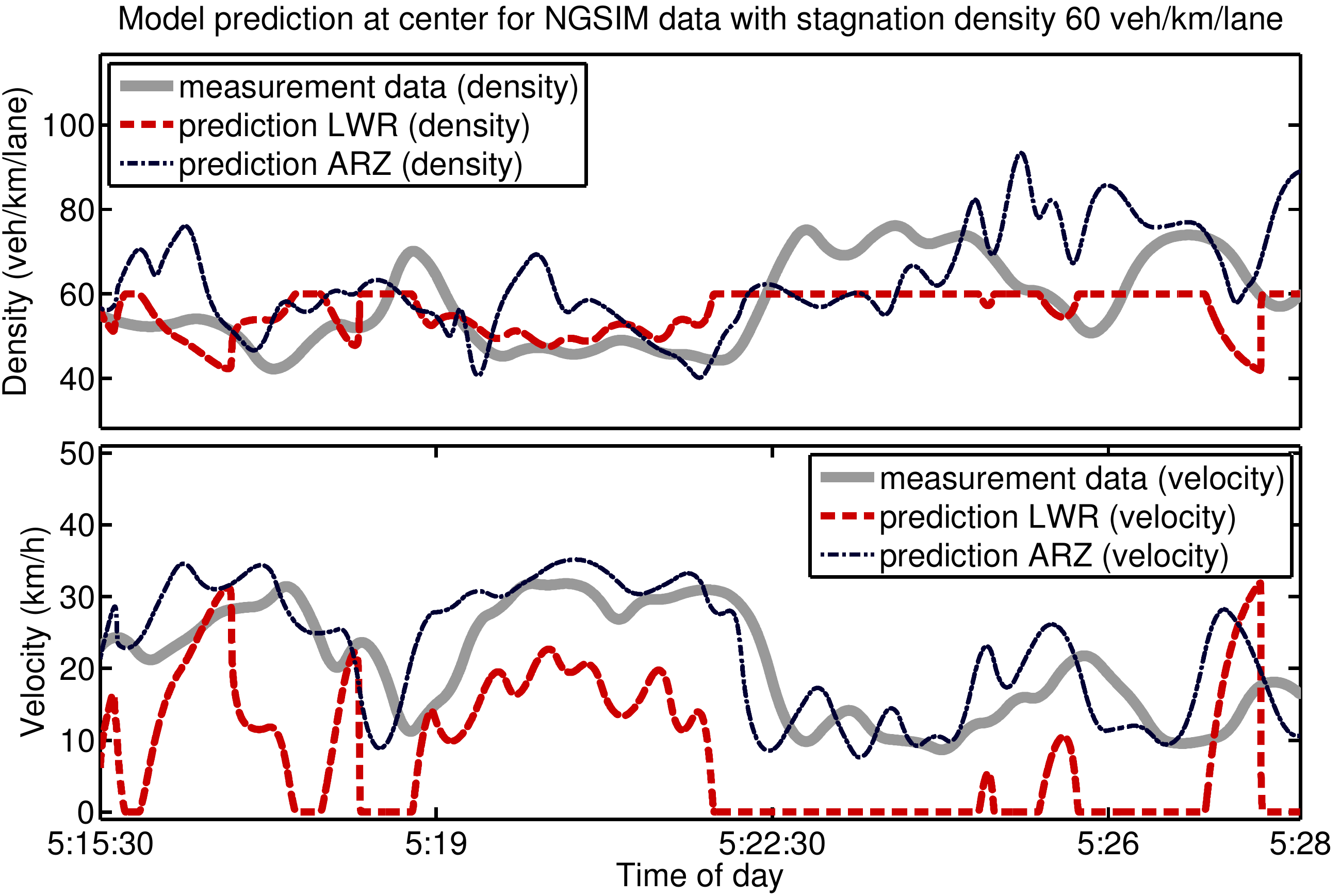}
\end{minipage}

\vspace{1em}
\begin{minipage}[b]{.412\textwidth}
\includegraphics[width=\textwidth]{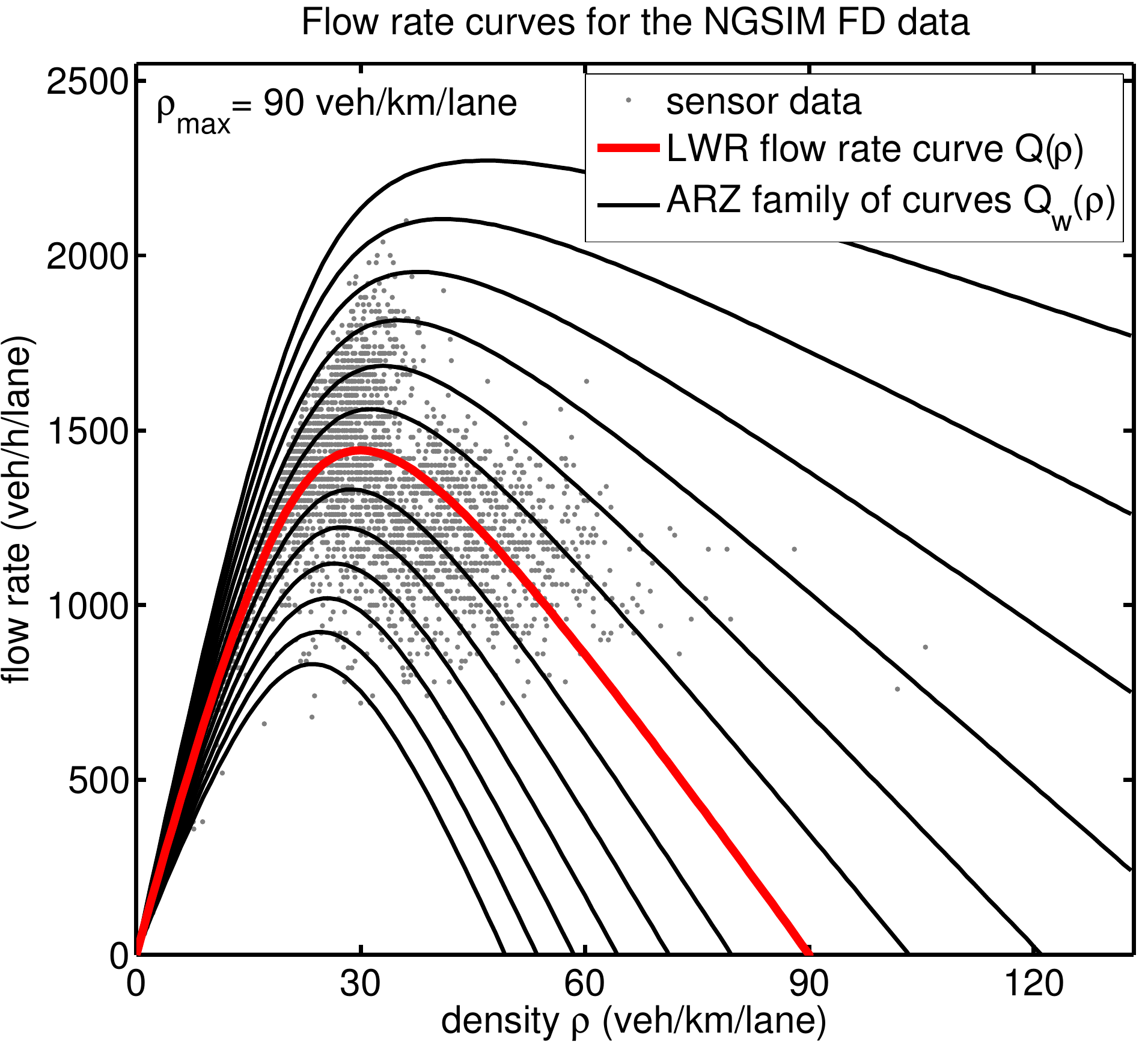}
\end{minipage}
\hfill
\begin{minipage}[b]{.56\textwidth}
\includegraphics[width=\textwidth]{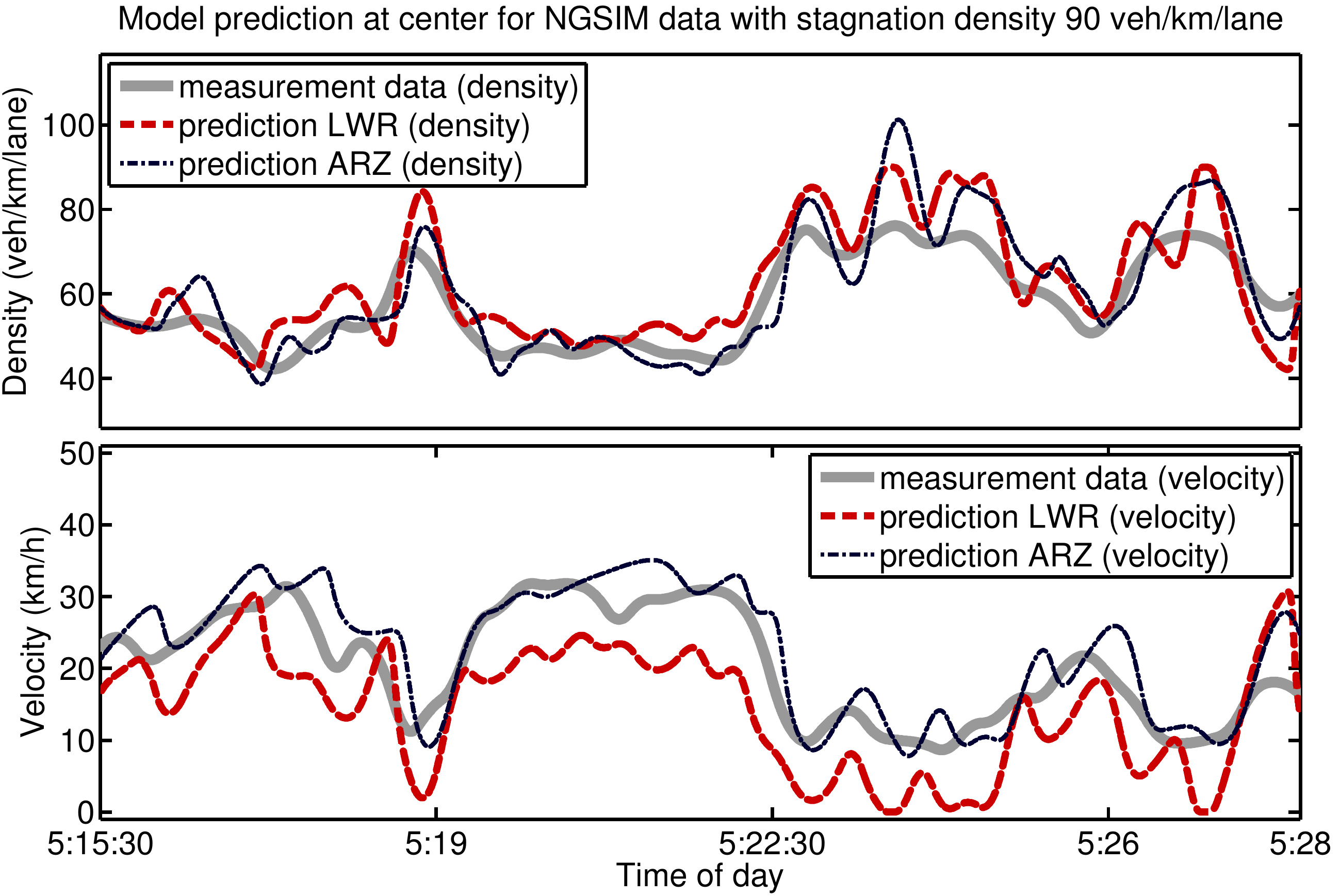}
\end{minipage}

\vspace{1em}
\begin{minipage}[b]{.412\textwidth}
\includegraphics[width=\textwidth]{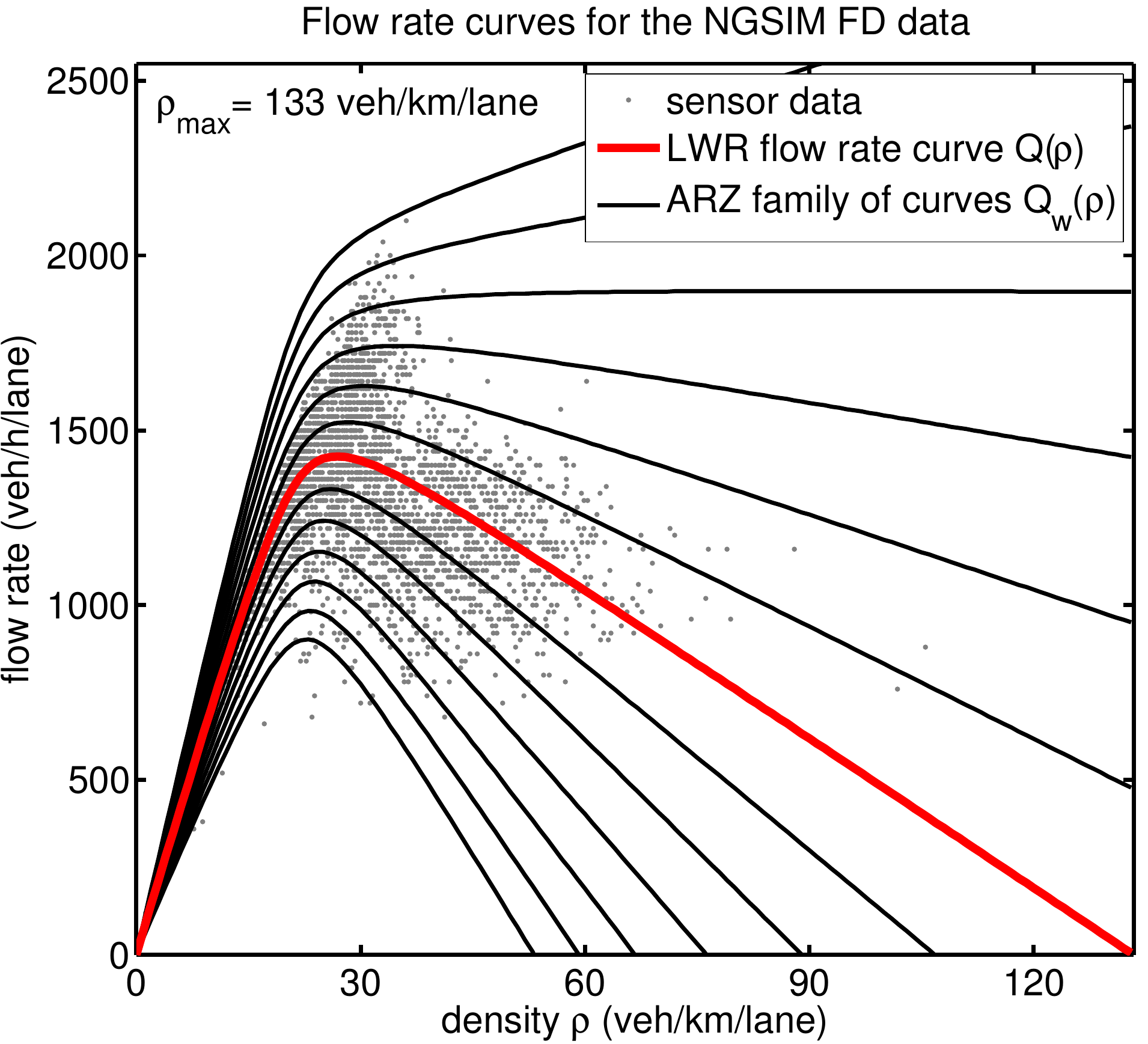}
\end{minipage}
\hfill
\begin{minipage}[b]{.56\textwidth}
\includegraphics[width=\textwidth]{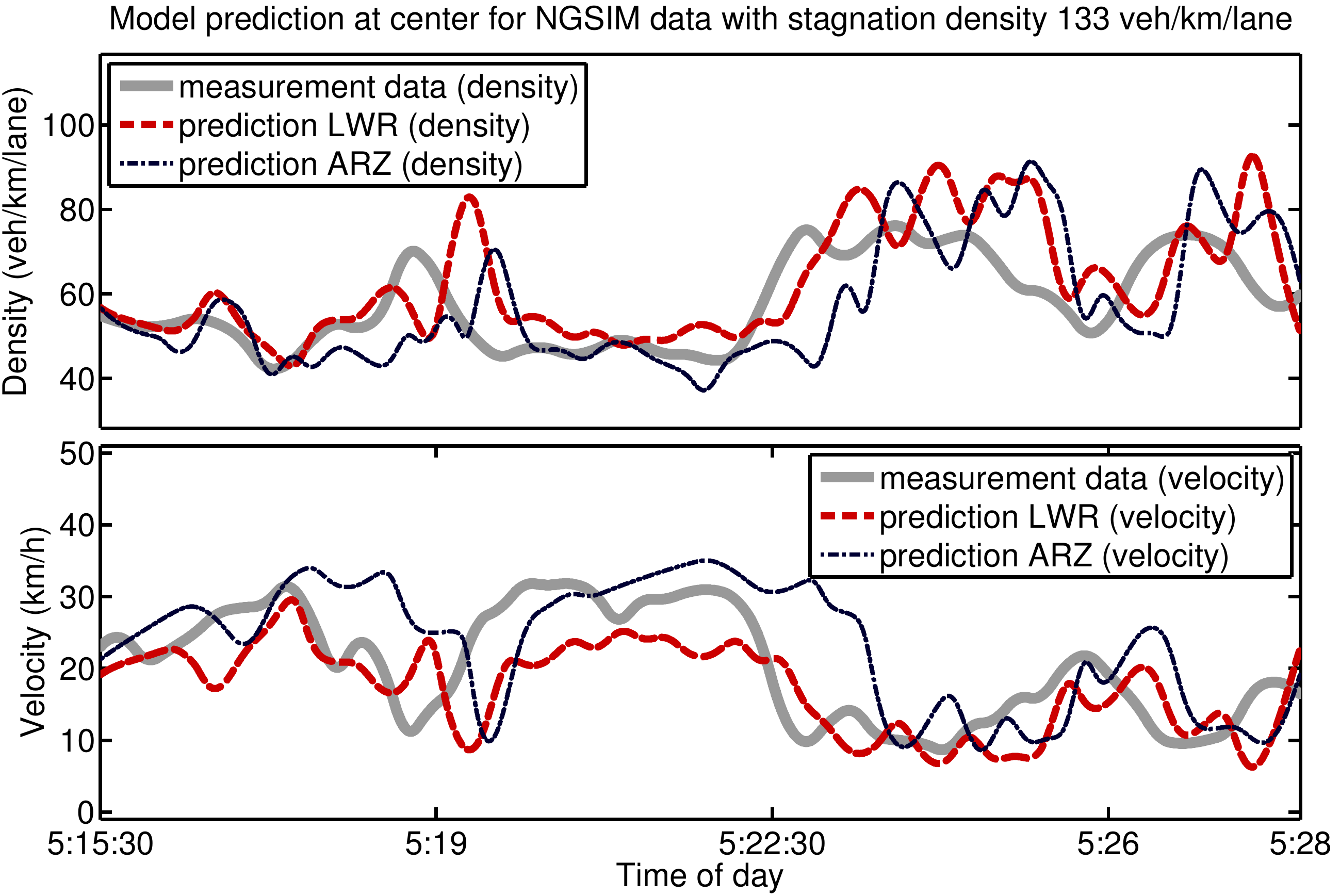}
\end{minipage}

\caption{Model predictions at a fixed position (mid-point of the domain) for the NGSIM 5:15--5:30 data set. The three rows correspond to three choices of $\rho_\text{max} \in \{60,90,133.33\}\, \text{veh}/\text{km}/\text{lane}$. The left column shows the flow rate curves for LWR (red) and ARZ (black), obtained via LSQ-fitting. The right column shows the temporal evolution of the predicted densities (top) and velocities (bottom) vs.\ the truth (gray).}
\label{fig:evolution_ngsim}
\end{figure}

\begin{figure}[p]
\begin{minipage}[b]{.412\textwidth}
\includegraphics[width=\textwidth]{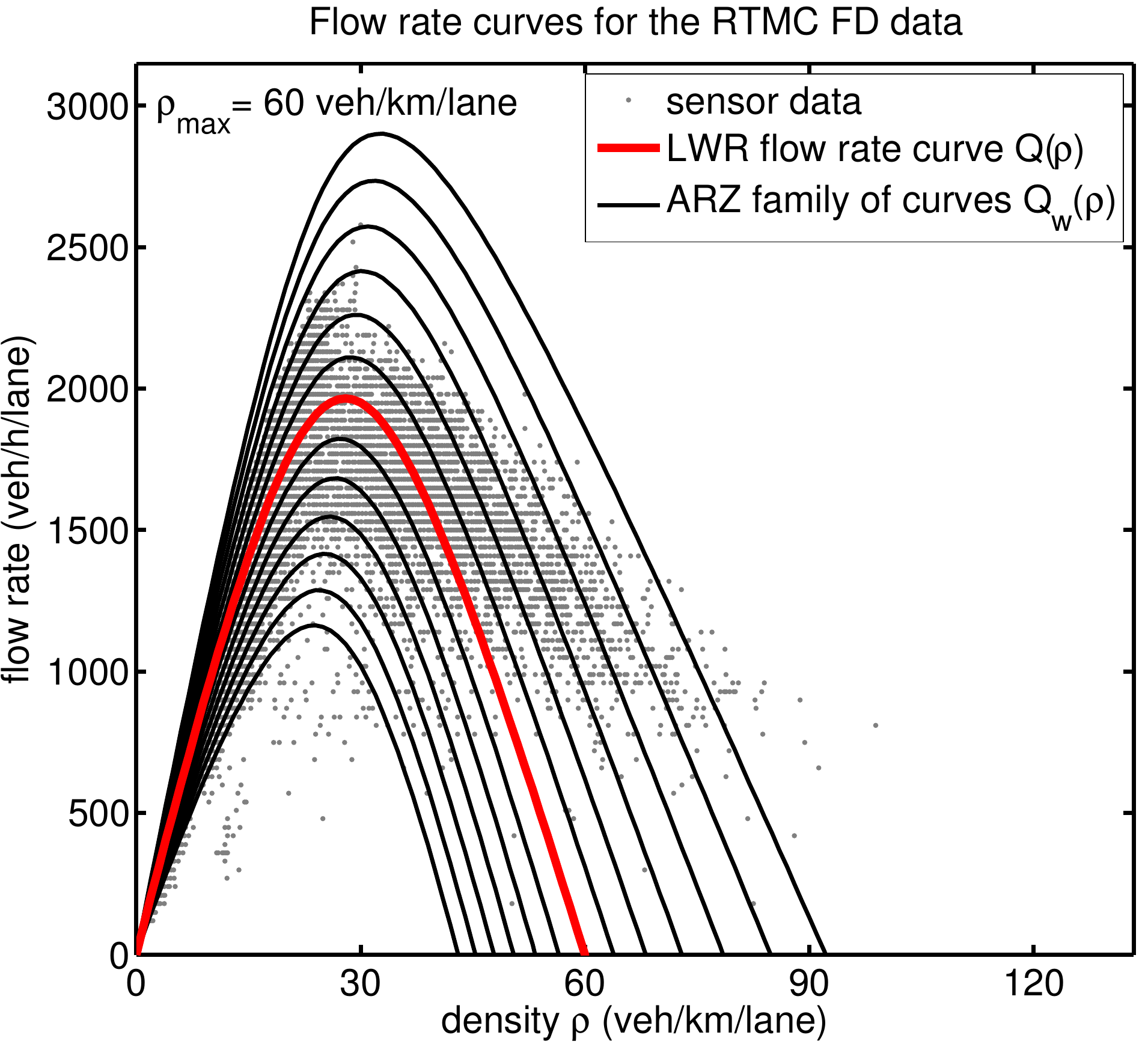}
\end{minipage}
\hfill
\begin{minipage}[b]{.56\textwidth}
\includegraphics[width=\textwidth]{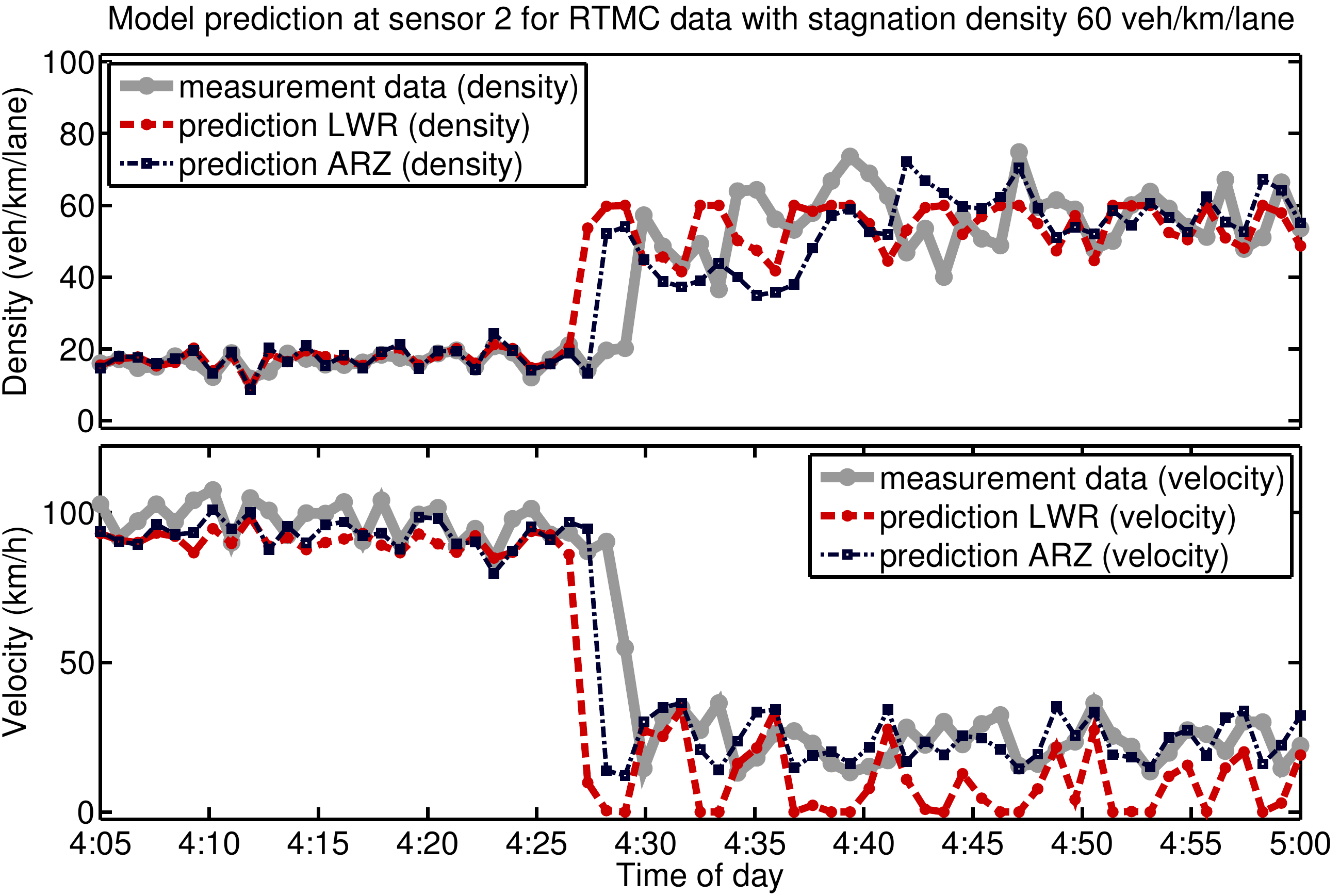}
\end{minipage}

\vspace{1em}
\begin{minipage}[b]{.412\textwidth}
\includegraphics[width=\textwidth]{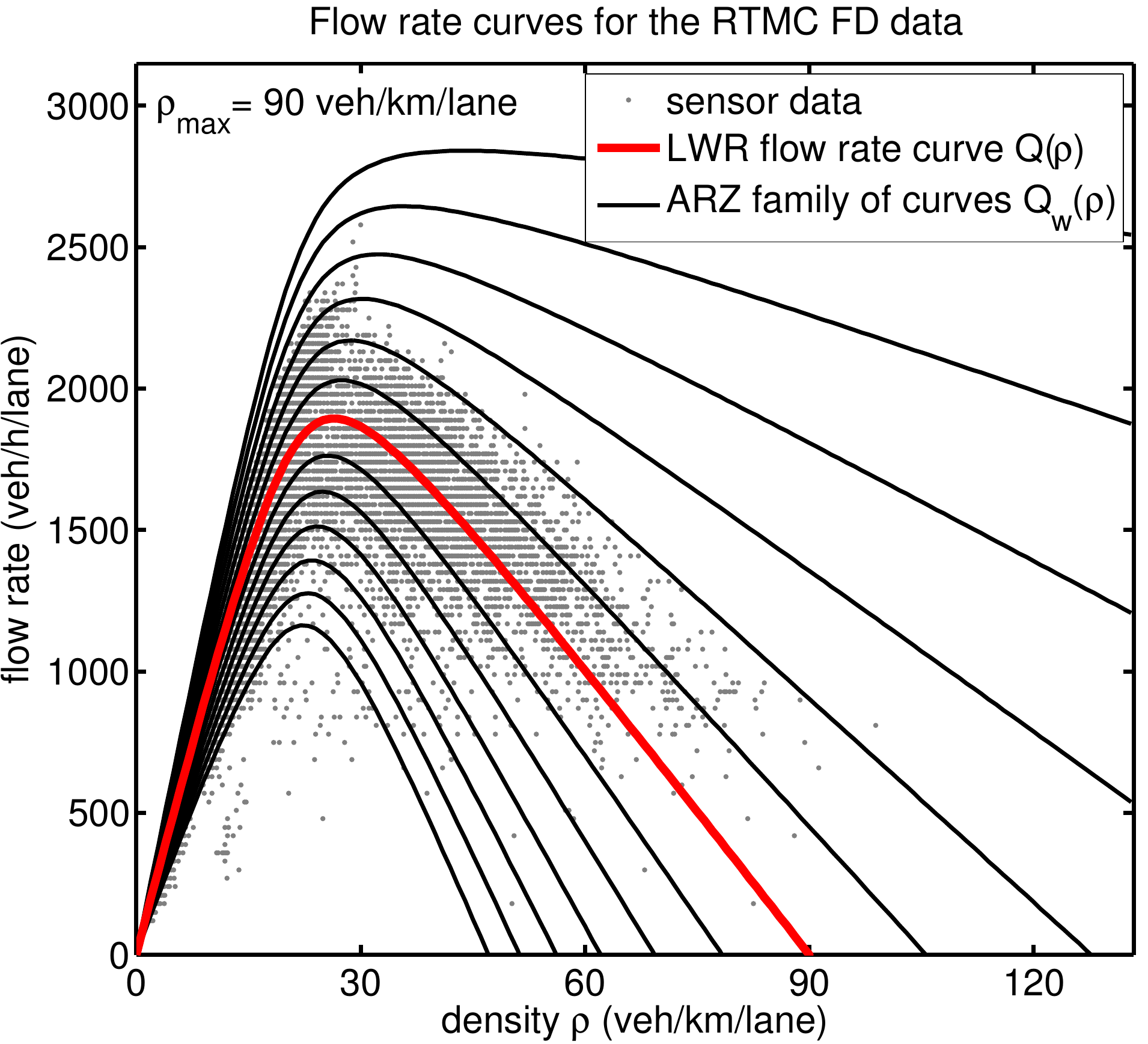}
\end{minipage}
\hfill
\begin{minipage}[b]{.56\textwidth}
\includegraphics[width=\textwidth]{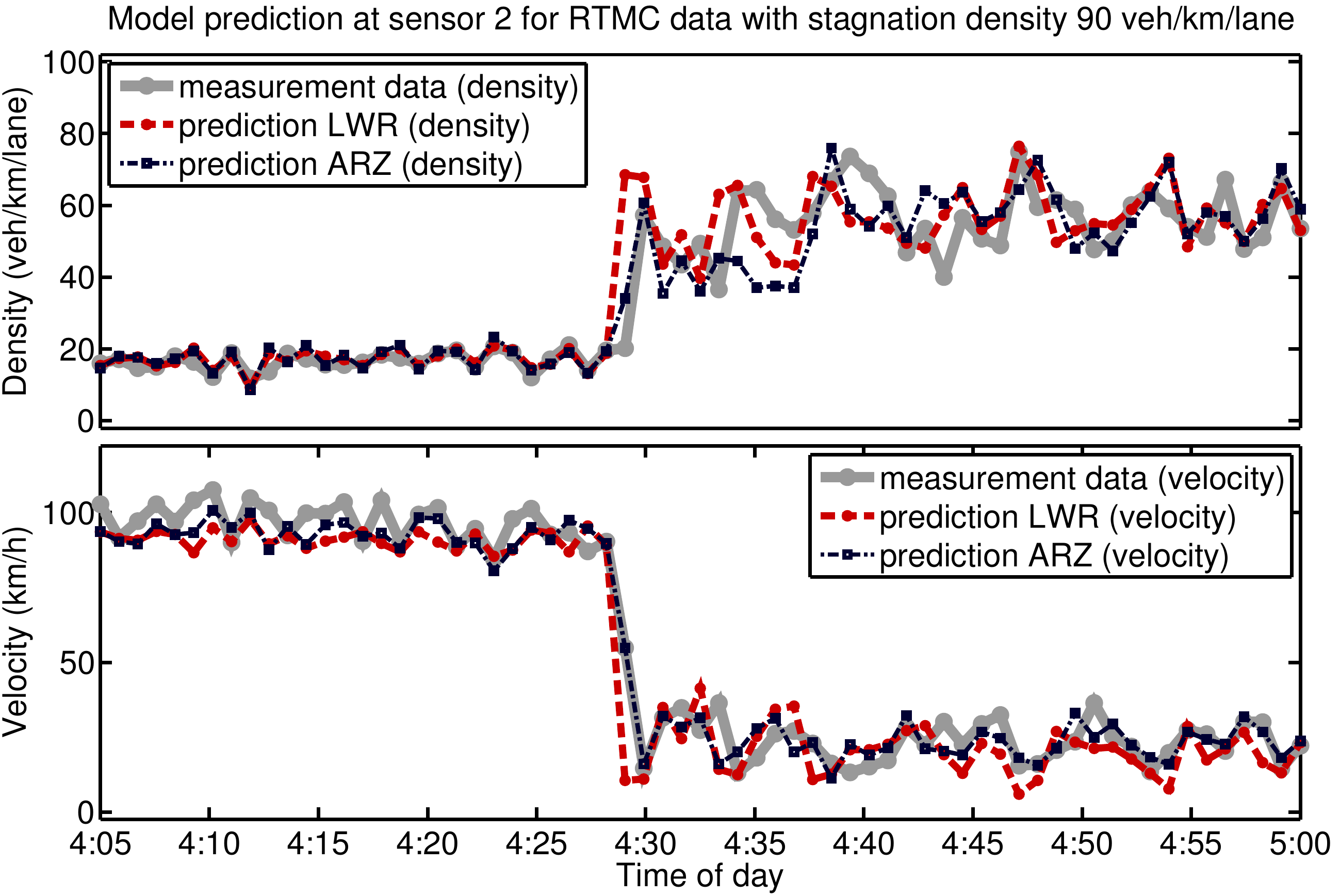}
\end{minipage}

\vspace{1em}
\begin{minipage}[b]{.412\textwidth}
\includegraphics[width=\textwidth]{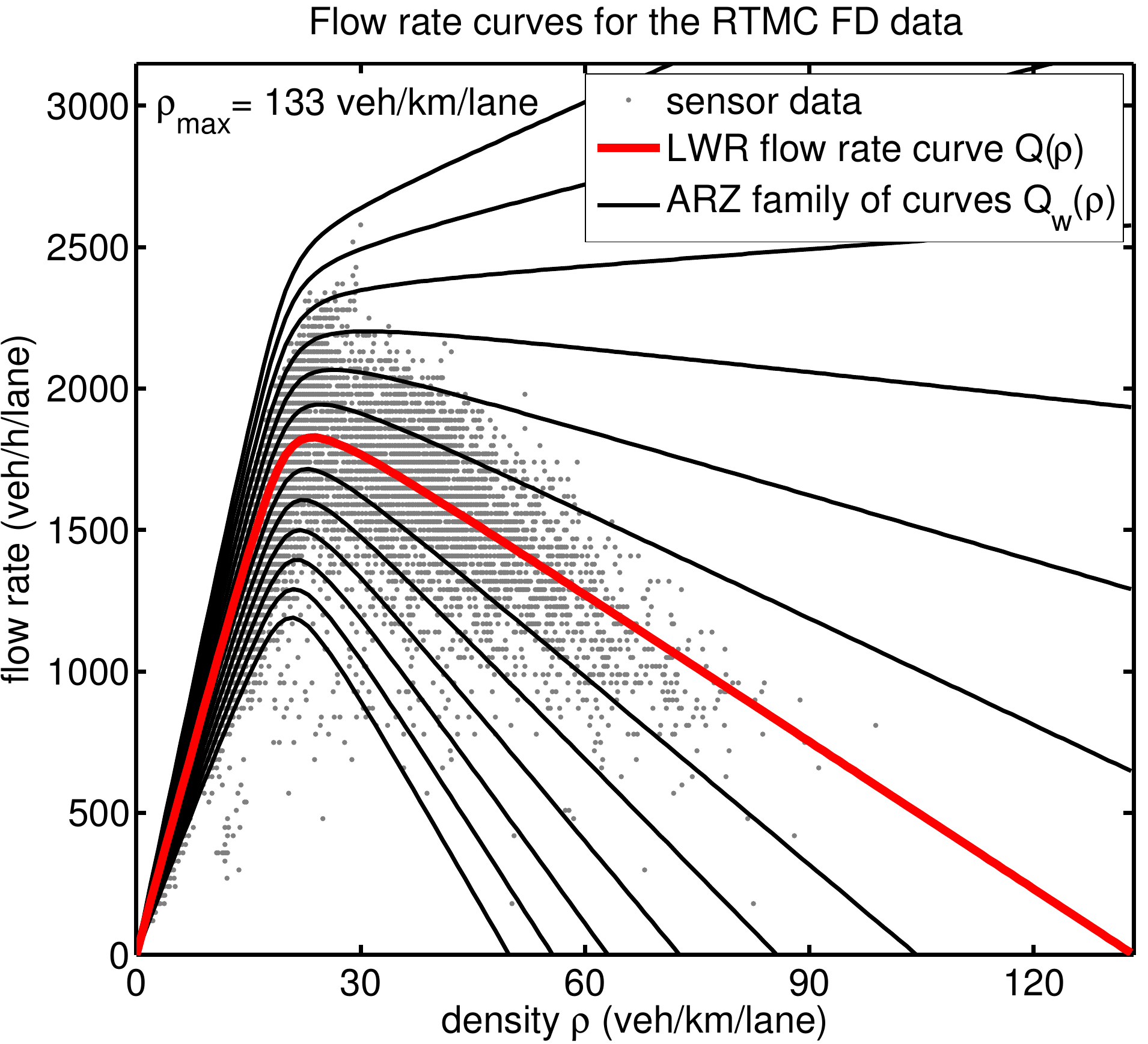}
\end{minipage}
\hfill
\begin{minipage}[b]{.56\textwidth}
\includegraphics[width=\textwidth]{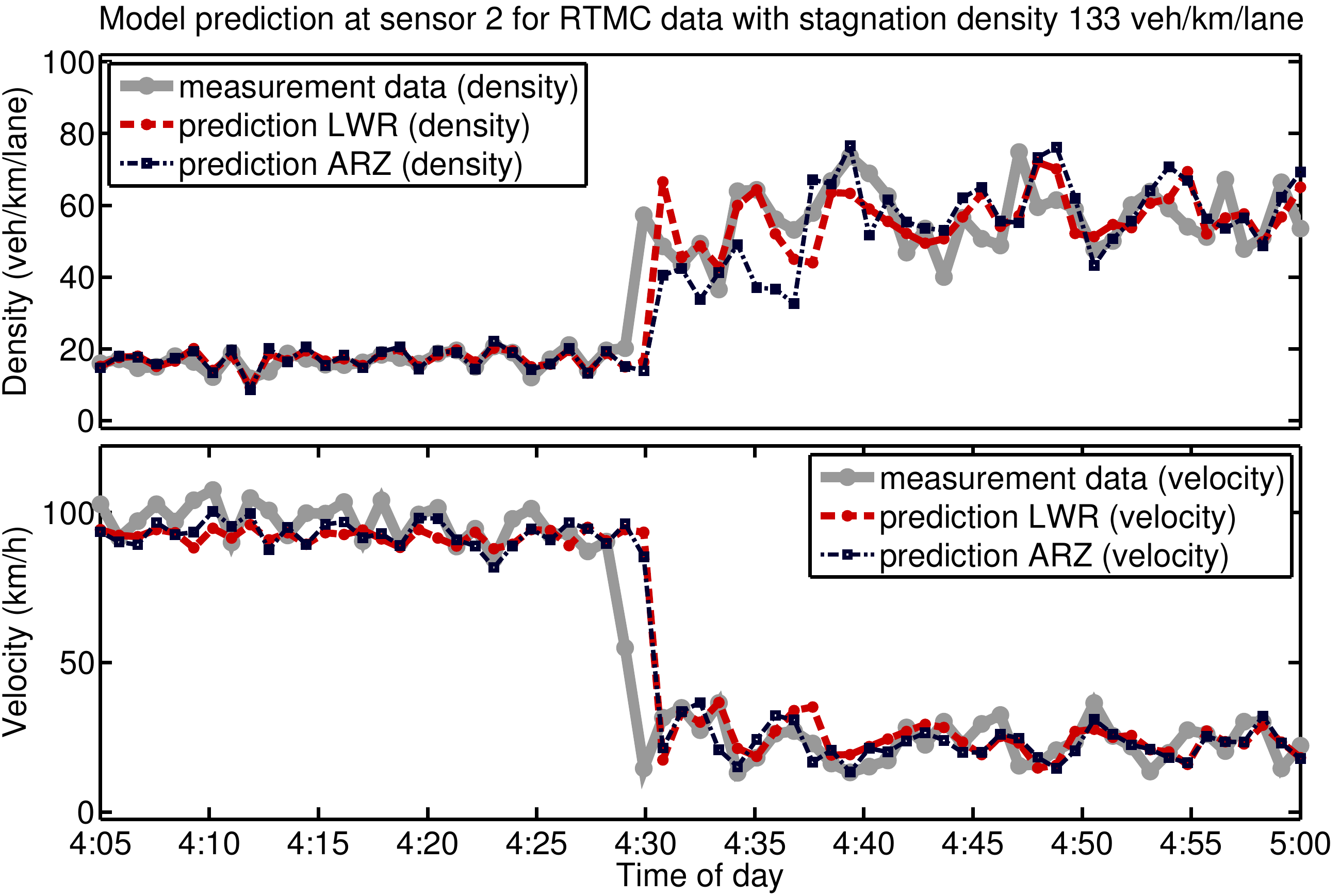}
\end{minipage}

\caption{Model predictions at a fixed position (sensor 2) for the RTMC data set. The three rows correspond to three choices of $\rho_\text{max} \in \{60,90,133.33\}\, \text{veh}/\text{km}/\text{lane}$. The left column shows the flow rate curves for LWR (red) and ARZ (black), obtained via LSQ-fitting. The right column shows the temporal evolution of the predicted densities (top) and velocities (bottom) vs.\ the truth (gray), for the rush hour on a typical day.}
\label{fig:evolution_rtmc}
\end{figure}

\vspace{1.5em}
\section{Results Part I: Wave Propagation Speed}
\label{sec:results_wave_speed}
We first focus on the speed at which information propagates. To that end, we consider the NGSIM data set with the largest congestion level (5:15--5:30), as well as a typical day (01/01/2003) in the RTMC data set on which a significant congestion level kicks in between 4pm and 5pm. In both cases we conduct the three-detector test described above, with the LWR and the ARZ model, and we evaluate the model predictions at a position $\bar{x}$ inside the study domain (the mid-point for NGSIM, and sensor 2 for RTMC). The resulting temporal profiles $\rho(\bar{x},t)$ and $u(\bar{x},t)$ are then compared with the evolution of the true data at the same position.

This test is conducted for three different choices of the stagnation density, \linebreak $\rho_\text{max} \in \{60,90,133.33\}\, \text{veh}/\text{km}/\text{lane}$, the latter being the choice used in \cite{FanSeibold2013}. The value of $\rho_\text{max}$ affects the shape of the flow rate curve \eqref{eq:flow_rate_vs_density_function}, that is obtained via LSQ-fitting to the FD data, as described in \S\ref{sec:model_creation}. A second effect is that boundary data prescribed in the LWR model is capped at $\rho_\text{max}$ (this happens to be of relevance only for $\rho_\text{max} = 60\, \text{veh}/\text{km}/\text{lane}$).

The results of this test are shown in Fig.~\ref{fig:evolution_ngsim} for the NGSIM data, and in Fig.~\ref{fig:evolution_rtmc} for the RTMC data. In each figure, the three rows correspond to the three choices of $\rho_\text{max}$. In each row, the left panel shows the FD data (gray dots), the LSQ-fitted LWR curve (red), and the induced family of ARZ curves (black). The right panel shows the temporal evolution of density (top) and velocity (bottom) as predicted by the LWR model (red) and the ARZ model (black), together with the true measurement data at the same position (gray).

First, we observe that generally, the ARZ model yields better predictions than the LWR model, in particular for the velocities. However, what is more important is the timing with which the models predict sudden changes in the traffic state. For instance, in Fig.~\ref{fig:evolution_ngsim} a plateau of high velocity is apparent between 5:19:30 and 5:22:30. The traffic models visibly reproduce this behavior; however, with $\rho_\text{max} = 60\, \text{veh}/\text{km}/\text{lane}$, transitions occur too early; in turn, with $\rho_\text{max} = 133.33\, \text{veh}/\text{km}/\text{lane}$, transitions occur too late. It is with $\rho_\text{max} = 90\, \text{veh}/\text{km}/\text{lane}$ that the timing of the features is right on target (particularly for the ARZ model).

That this is not a fluke becomes apparent when inspecting the results in Fig.~\ref{fig:evolution_rtmc}. The sudden transition from free flow to congestion at 4:29pm is reproduced almost perfectly if $\rho_\text{max} = 90\, \text{veh}/\text{km}/\text{lane}$. In contrast, the models with $\rho_\text{max} = 60\, \text{veh}/\text{km}/\text{lane}$ predict the event 2 minutes too early; and with $\rho_\text{max} = 133.33\, \text{veh}/\text{km}/\text{lane}$ the event is predicted 1 minute too late.

These observations can be explained as follows. Since all examples involve congestion, information travels backwards on the road. A feature (such as the sudden jump in density visible in Fig.~\ref{fig:evolution_rtmc}) first appears at the right boundary $x_\text{R}$, and then travels into the domain. The feature is recorded when it reaches the evaluation point $\bar{x}$. Therefore, the models have the right timing if they propagate information backwards at the correct speed. In turn, the models' predictions are too early (late) if information propagates too fast (slowly). As described in \S\ref{subsec:wave_speed}, the speed of information propagation is given by the slope of the curves $Q(\rho)$ and $Q_w(\rho)$, shown in the left panels of Figs.~\ref{fig:evolution_ngsim} and~\ref{fig:evolution_rtmc}. The fact that the smaller $\rho_\text{max}$, the steeper the decrease of these curves, explains the observations.

An intriguing aspect of these results is that the stagnation density that yields correct wave propagation is unexpectedly low. In turn, stagnation densities that have been suggested in the literature (cf.~\cite{Helbing2001}) of $\rho_\text{max} = 120\, \text{veh}/\text{km}/\text{lane}$ and above generate systematically too slow information propagation.

\begin{figure}
\begin{minipage}[b]{.485\textwidth}
\includegraphics[width=\textwidth]{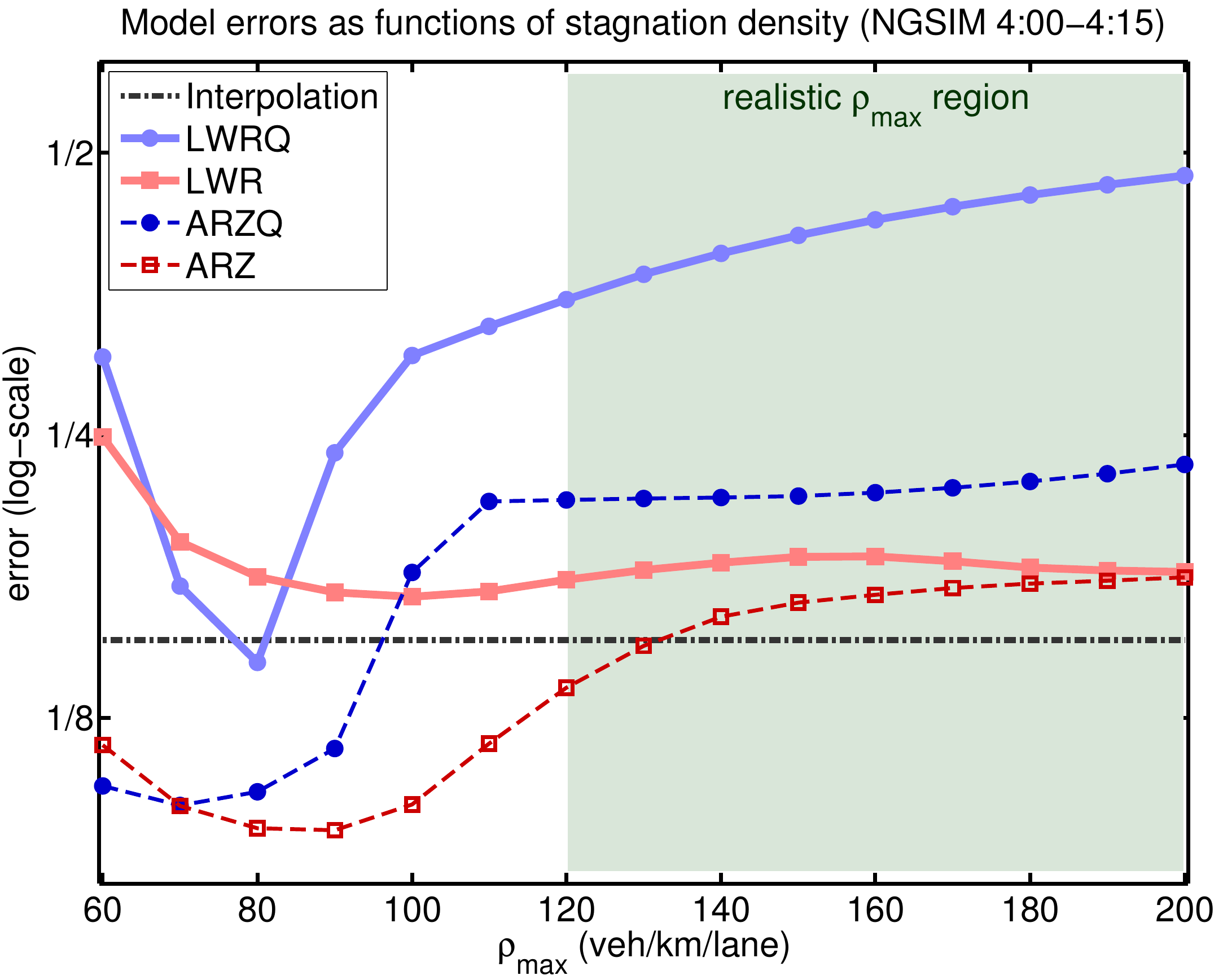}
\end{minipage}
\hfill
\begin{minipage}[b]{.485\textwidth}
\includegraphics[width=\textwidth]{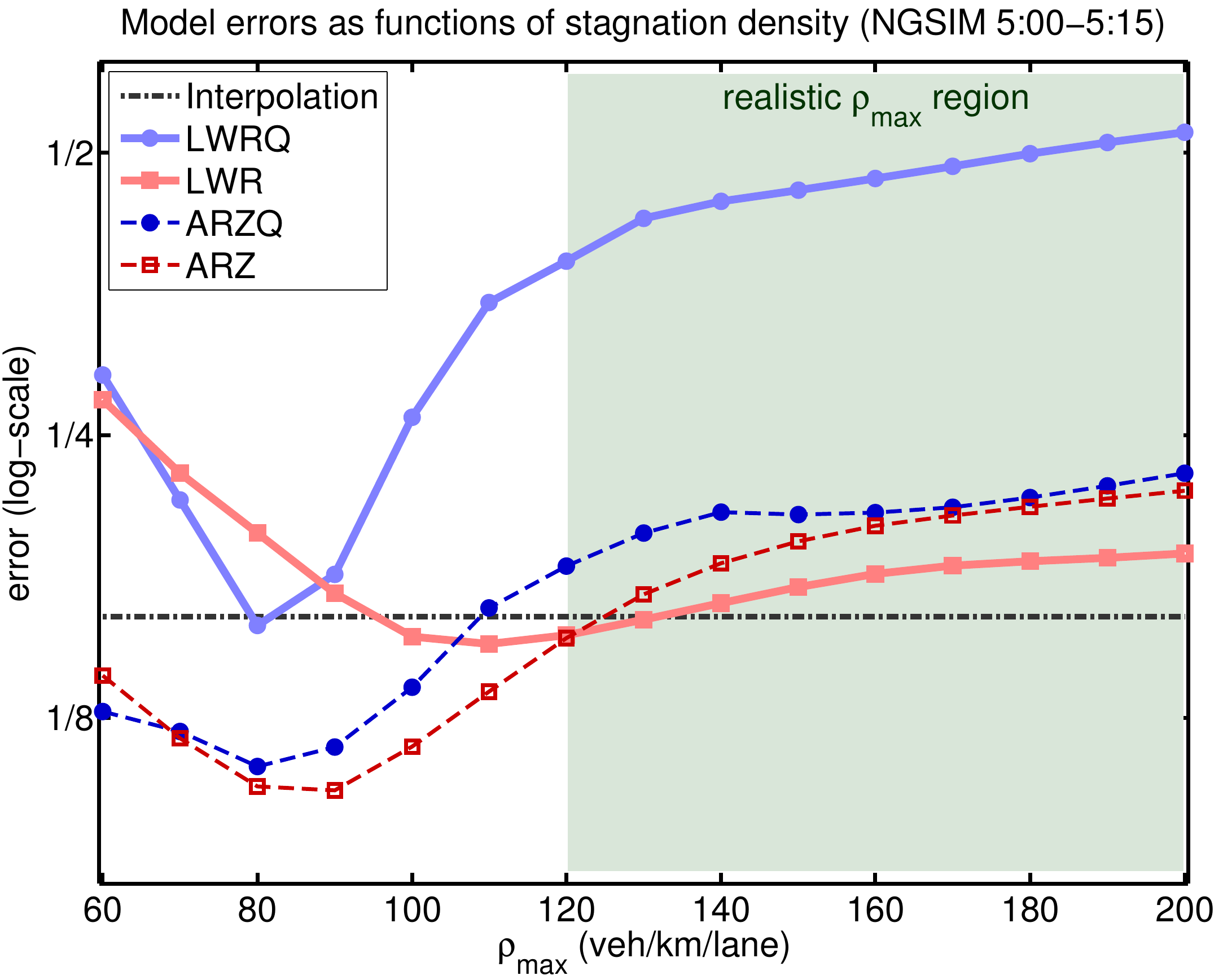}
\end{minipage}

\vspace{.4em}
\begin{minipage}[b]{.485\textwidth}
\includegraphics[width=\textwidth]{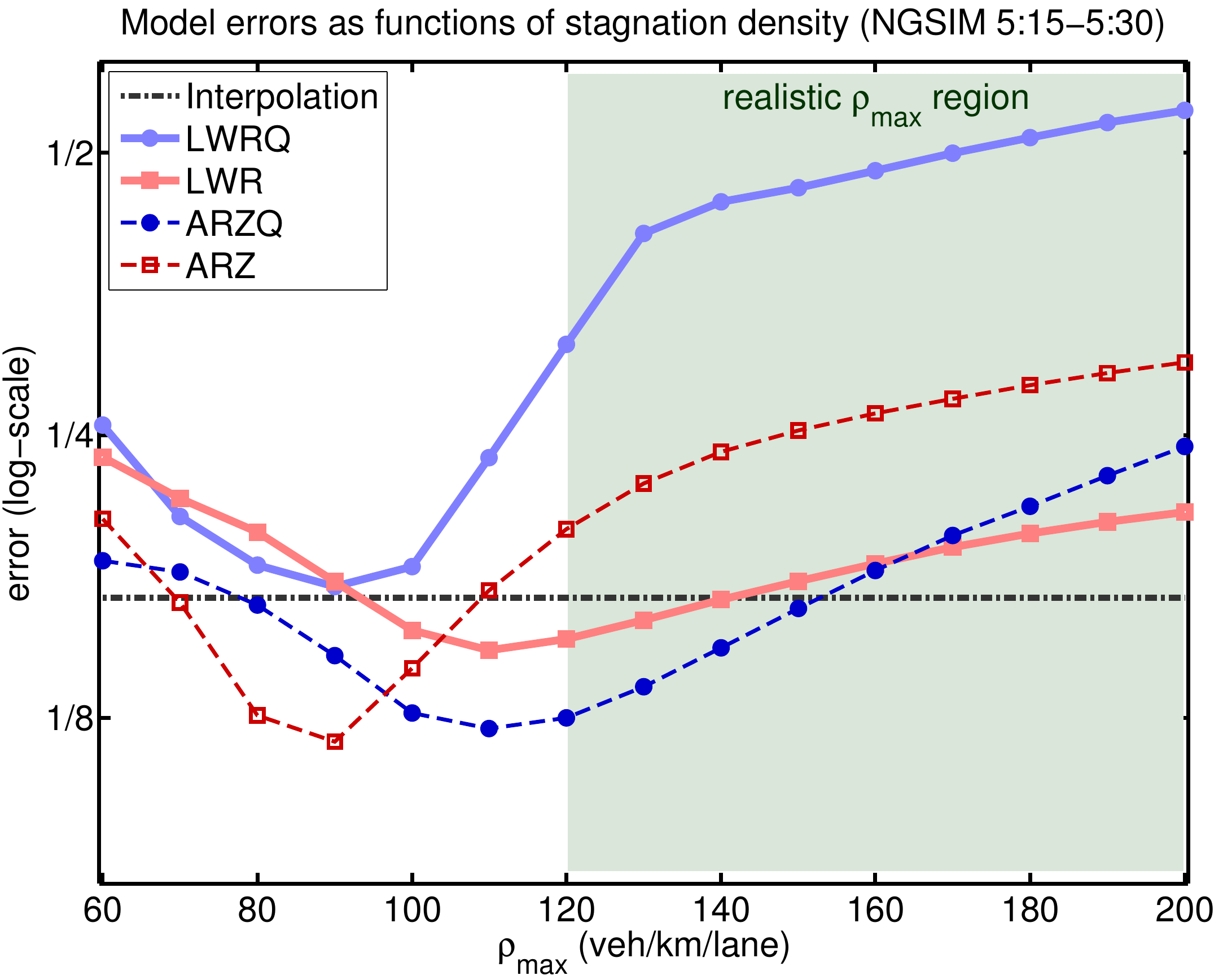}
\end{minipage}
\hfill
\begin{minipage}[b]{.485\textwidth}
\includegraphics[width=\textwidth]{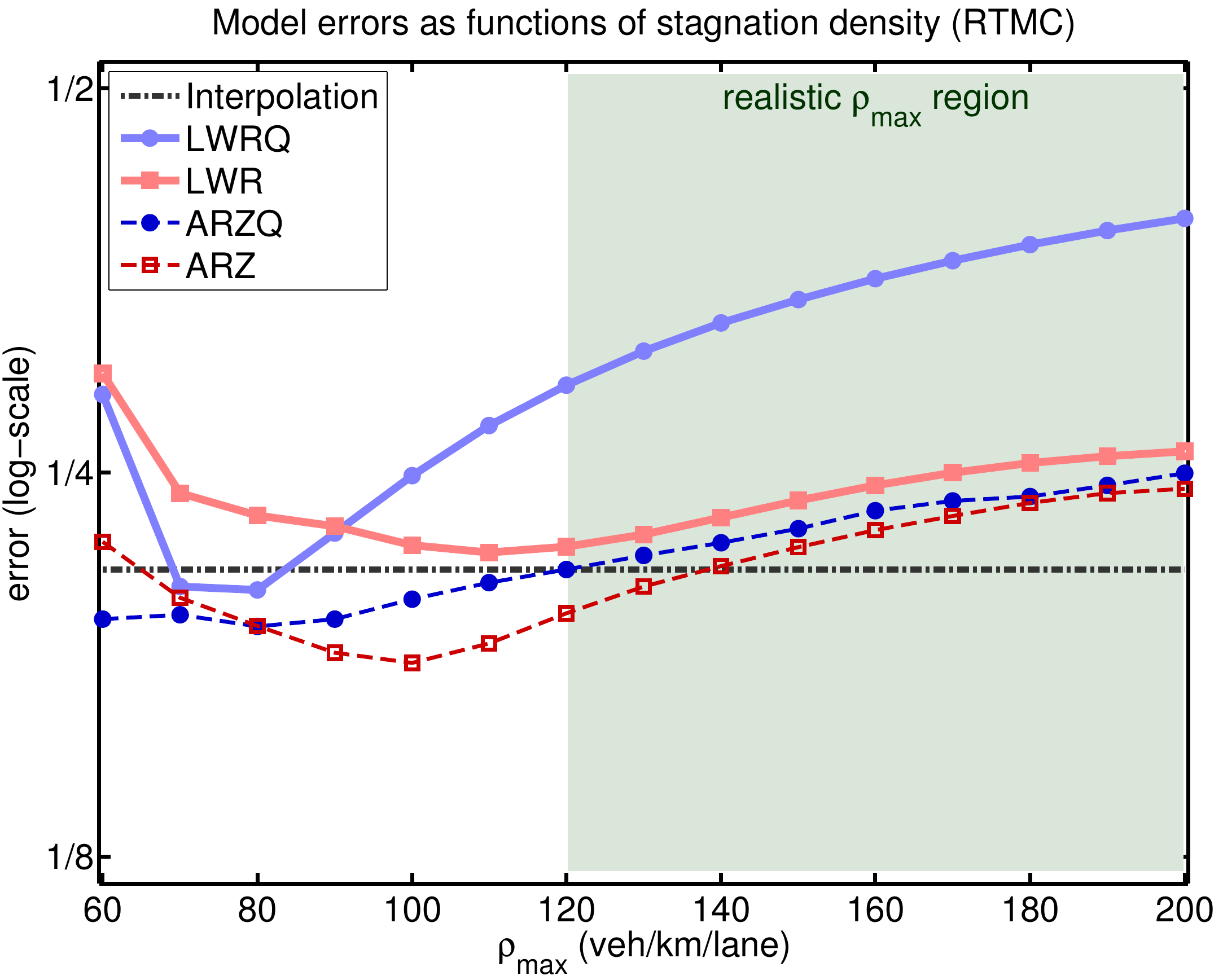}
\end{minipage}

\vspace{-.5em}
\caption{Errors of various models as functions of the stagnation density $\rho_\text{max}$. The panels show the space-time errors \eqref{eq:error_xt} for the three NGSIM data sets (top-left, top-right, and bottom-left), and the error \eqref{eq:error_days_t} for the RTMC data (bottom-right), averaged over time and all congested days. The four traffic models, LWRQ (light blue), LWR (light red), ARZQ (dark blue dashed), and ARZ (dark red dashed) are considered, as well as an Interpolation predictor (dark dash-dotted) that simply interpolates the traffic state from the boundaries.}
\label{fig:model_accuracy_rhomax}
\end{figure}

\vspace{1.5em}
\section{Results Part II: Model Accuracy as a Function of the Stagnation Density}
\label{sec:results_model_accuracy}
While the investigation in \S\ref{sec:results_wave_speed} focuses on specific features in the time evolution of the model predictions, we now study how the fully averaged model errors behave as functions of $\rho_\text{max}$. As described in \S\ref{subsec:error_measures}, the total error for an NGSIM data set is obtained by averaging the deviation of the predicted state from the data with respect to space and time; and the total error for the RTMC data is obtained by averaging over time and the 43 study days. This averaging is important, because it removes noise and thus reveals the overall trends in the predictive qualities of the models.

In this model comparison we consider the four traffic models defined in \S\ref{sec:model_creation}: LWR, LWRQ, ARZ, and ARZQ. Moreover, we include a reference predictor denoted ``Interpolation'', which is defined as follows. At any instance in time, the traffic state at a position $x\in [x_\text{L},x_\text{R}]$ is given by the linear interpolant of the boundary data, i.e., $\rho(x,t) = \rho(x_\text{L},t) + (\rho(x_\text{R},t)-\rho(x_\text{L},t)) \frac{x-x_\text{L}}{x_\text{R}-x_\text{L}}$, and analogous for $u$. Clearly, the Interpolation predictor is not a traffic model; and in particular, it is independent of $\rho_\text{max}$.

The results of this study are shown in Fig.~\ref{fig:model_accuracy_rhomax}. The first three panels represent the three NGSIM data sets, and the bottom-right panel corresponds to the RTMC data. Stagnation densities in the interval $\rho_\text{max}\in [60,200]\; \text{veh}/\text{km}/\text{lane}$ are considered. The range of realistic values that is suggested in \cite{Helbing2001} is marked via a gray-green background. The model errors are given in log-scale, i.e., the distance between two ticks on the vertical axis corresponds to a factor of 2 in the model errors.

One can see that in each test, each of the four traffic models possess an optimal stagnation density $\rho_\text{max}^\text{opt}$ for which the model error is minimized. In all cases, this $\rho_\text{max}^\text{opt}$ lies below the range of supposedly realistic values, thus indicating that the observations in \S\ref{sec:results_wave_speed} hold more generally: with $\rho_\text{max} \ge 120\, \text{veh}/\text{km}/\text{lane}$, information propagates too slowly in traffic models; and choosing smaller values improves the model accuracy.

Table~\ref{tab:models_optimal_rhomax} collects the errors for each model (in each of the four test cases) when conducted with the optimal stagnation density. One can see that, with the exception of the ARZQ model, the choice of $\rho_\text{max}^\text{opt}$ is only mildly dependent on the particular test case. For each test case, the actual error values of the four models are reported, and (in parentheses) the relative excess error, $E/E_\text{ARZ}-1$, with respect to the best model (ARZ), measured in percent. The results confirm that with optimal stagnation density, the second-order models yield significant improvements over the first-order models, whose model errors are up to 77\% larger.

\begin{table}
\begin{tabular}{|l||
@{}r@{\hspace{.4em}}|r@{\hspace{.5em}}r@{\hspace{.3em}}|
@{}r@{\hspace{.4em}}|r@{\hspace{.5em}}r@{\hspace{.3em}}|
@{}r@{\hspace{.4em}}|r@{\hspace{.5em}}r@{\hspace{.3em}}|
@{}r@{\hspace{.4em}}|r|}
\hline
& \multicolumn{3}{c|}{LWRQ\rule{0em}{1.04em}} & \multicolumn{3}{c|}{LWR}
& \multicolumn{3}{c|}{ARZQ} & \multicolumn{2}{c|}{ARZ} \\
\cline{2-12}
Data set\rule{0em}{1.04em} &
\hspace{.2em}$\rho_\text{max}^\text{opt}$\hspace{-.3em} & \multicolumn{2}{c|}{error} &
\hspace{.2em}$\rho_\text{max}^\text{opt}$\hspace{-.3em} & \multicolumn{2}{c|}{error} &
\hspace{.2em}$\rho_\text{max}^\text{opt}$\hspace{-.3em} & \multicolumn{2}{c|}{error} &
\hspace{.2em}$\rho_\text{max}^\text{opt}$\hspace{-.3em} & \multicolumn{1}{c|}{error} \\
\hline\hline
NGSIM~~4:00\rule{0em}{1.04em}
            &80&0.143&(+51\%)&100&0.168&(+77\%)& 70&0.101&(+6\%)& 90&0.095\\
NGSIM~~5:00 &80&0.157&(+50\%)&110&0.150&(+43\%)& 80&0.111&(+6\%)& 90&0.105\\
NGSIM~~5:15 &90&0.173&(+46\%)&110&0.148&(+25\%)&110&0.122&(+3\%)& 90&0.118\\
\hline
RTMC\rule{0em}{1.04em}
            &80&0.202&(+14\%)&110&0.216&(+22\%)& 80&0.189&(+7\%)&100&0.177\\
\hline
\end{tabular}
\vspace{.6em}
\caption{Comparison of traffic models for the NGSIM data sets and the RTMC data on congested days. Based on the study in Fig.~\ref{fig:model_accuracy_rhomax}, for each data set and model, the optimal stagnation density is determined, and the corresponding model error is reported. In parentheses shown are the relative excess errors of the models with respect to the most accurate model in the particular test (which is the ARZ model).}
\label{tab:models_optimal_rhomax}
\end{table}

\begin{table}
\begin{tabular}{|l||
r@{\hspace{.8em}}r@{\hspace{.8em}}|
r@{\hspace{.8em}}r@{\hspace{.8em}}|
r@{\hspace{.8em}}r@{\hspace{.8em}}|
r@{\hspace{.8em}}r@{\hspace{.8em}}|}
\hline
& \multicolumn{2}{c|}{LWRQ\rule{0em}{1.04em}} & \multicolumn{2}{c|}{LWR}
& \multicolumn{2}{c|}{ARZQ} & \multicolumn{2}{c|}{ARZ} \\
\cline{2-9}
Data set\rule{0em}{1.10em} &
\hspace{-.4em}$\frac{E_{133.33}}{E_\text{opt}}\!\!-\!\!1$\hspace{-.4em} &
$\frac{E_{133.33}}{E_{100}}\!\!-\!\!1$\hspace{-.6em} &
\hspace{-.4em}$\frac{E_{133.33}}{E_\text{opt}}\!\!-\!\!1$\hspace{-.4em} &
$\frac{E_{133.33}}{E_{100}}\!\!-\!\!1$\hspace{-.6em} &
\hspace{-.4em}$\frac{E_{133.33}}{E_\text{opt}}\!\!-\!\!1$\hspace{-.4em} &
$\frac{E_{133.33}}{E_{100}}\!\!-\!\!1$\hspace{-.6em} &
\hspace{-.3em}$\frac{E_{133.33}}{E_\text{opt}}\!\!-\!\!1$\hspace{-.4em} &
$\frac{E_{133.33}}{E_{100}}\!\!-\!\!1$\hspace{-.6em} \\
\hline\hline
NGSIM~~4:00\rule{0em}{1.04em}
            &+164\%& +24\%  &  +7\%&  +7\%  &+112\%& +20\%  & +62\%& +52\% \\
NGSIM~~5:00 &+176\%& +65\%  &  +8\%&  +6\%  & +81\%& +49\%  & +67\%& +50\% \\
NGSIM~~5:15 &+150\%&+138\%  & +10\%&  +4\%  & +14\%& +10\%  & +94\%& +62\% \\
\hline
RTMC\rule{0em}{1.04em}
            & +57\%& +28\%  &  +4\%&  +3\%  & +13\%&  +7\%  & +16\%& +16\% \\
\hline
\end{tabular}
\vspace{.6em}
\caption{Excess errors for the traffic models with $\rho_\text{max} = 133.33\,\text{veh}/\text{km}/\text{lane}$, relative to the same model with $\rho_\text{max} = \rho_\text{max}^\text{opt}$ (left number) and relative to the same model with $\rho_\text{max} = 100\,\text{veh}/\text{km}/\text{lane}$ (right number). The latter turns out to be a good general choice if $\rho_\text{max}^\text{opt}$ is not known.}
\label{tab:models_improvement}
\end{table}

Table~\ref{tab:models_improvement} shows the relative excess error of each model with $\rho_\text{max} = 133.33\, \text{veh}/\text{km}/\text{lane}$ (as used in \cite{FanSeibold2013}), relative to the error obtained with the same model when using \\
a) $\rho_\text{max} = \rho_\text{max}^\text{opt}$, i.e., the best one can do when choosing $\rho_\text{max}$ freely; and \\
b) $\rho_\text{max} = 100\, \text{veh}/\text{km}/\text{lane}$, which is a good overall choice if one does not know $\rho_\text{max}^\text{opt}$. \\[.2em]
The improvements obtained by choosing the optimal stagnation density are quite significant for the models LWRQ, ARZQ, and ARZ (up to a factor of 2.8 for LWRQ). In turn, the LWR model is much less sensitive to the choice of $\rho_\text{max}$. The choice $\rho_\text{max} = 100\, \text{veh}/\text{km}/\text{lane}$ never yields worse results than $\rho_\text{max} = 133.33\, \text{veh}/\text{km}/\text{lane}$, and in some cases it improves the model accuracy tremendously. This is particularly apparent for the overall best model, ARZ. Further insights obtained from this study are:
\begin{enumerate}[1)]
\item The LWRQ model (light blue) is not a good choice; it hardly ever yields better predictions than the reference Interpolation predictor (dark dash-dotted), see Fig.~\ref{fig:model_accuracy_rhomax}.
\item The second-order models ARZ (dark red dashed) and ARZQ (dark blue dashed) yield better predictions than their first-order counterparts LWR (light red) and LWRQ. An exception from this rule is ARZ vs.~LWR for a high congestion level and when choosing $\rho_\text{max} > 100\, \text{veh}/\text{km}/\text{lane}$. The reason is that the ARZ model possesses an unrealistic range of densities at which the velocity vanishes, as transparent in the left panels of Figs.~\ref{fig:evolution_ngsim} and~\ref{fig:evolution_rtmc}. In reality, the curves $Q_w(\rho)$ should all vanish around the same density, since different empty road velocities should not affect the density at which drivers come to a halt.
\item The second-order models are significantly more accurate than the first-order models if the stagnation density is chosen optimally (see Table~\ref{tab:models_optimal_rhomax}) or nearly optimally (see Table~\ref{tab:models_improvement}).
\item Considering that the Interpolation predictor is not a traffic model, it yields surprisingly good results. The reason is that it uses a total of four pieces of data to make a prediction ($\rho$ and $u$ at each boundary), while the traffic models usually incorporate only one or two pieces of boundary data. Moreover, since interpolation reproduces information propagation incorrectly (it is instantaneous), one can expect that it perform worse on longer road segments. Finally, even in the test cases here, if $\rho_\text{max}$ is chosen well, the second-order models do outperform the Interpolation predictor.
\end{enumerate}

\vspace{1.5em}
\section{Conclusions and Outlook}
\label{sec:conclusions_and_outlook}
The results of this study reveal very clearly that in data-fitted traffic models, the choice of the stagnation density $\rho_\text{max}$ has a significant effect on the predictive quality of the models (in congested traffic). The dependence of the overall model errors on $\rho_\text{max}$, investigated in \S\ref{sec:results_model_accuracy}, reveals that choosing $\rho_\text{max}\in [90,100]\; \text{veh}/\text{km}/\text{lane}$ yields better models than when choosing $\rho_\text{max}\in [120,200]\; \text{veh}/\text{km}/\text{lane}$, as suggested in the literature \cite{Helbing2001}. In fact, in many cases the improvement is tremendous: the errors decrease by factors of 2 or more.

We also understand the main reason for this behavior. As the investigation in \S\ref{sec:results_wave_speed} shows, it is the propagation speed of information backwards on the road that largely is responsible for the predictive quality of the models. With $\rho_\text{max} = 60\, \text{veh}/\text{km}/\text{lane}$, information travels faster in the models than it does in reality; in turn, with $\rho_\text{max} = 133.33\, \text{veh}/\text{km}/\text{lane}$, information travels too slowly in the models. When choosing $\rho_\text{max} \approx 90\, \text{veh}/\text{km}/\text{lane}$, the wave propagation speed that the models possess matches with reality.

As a general conclusion from this study, we recommend to choose $\rho_\text{max} = 100\,\text{veh}/\text{km}/\text{lane}$ in data-fitted models of the type presented here. At the same time, this recommendation should inspire further thoughts about the classes of fundamental diagram functions, $Q(\rho)$, that one considers. In reality, when traffic comes to a complete stop, vehicles tend to leave less than a full vehicle length of space in between them. Hence, one should strive for models that a) reproduce correct information propagation speeds in congested, but moving, traffic, and b) possess stagnation densities of $\rho_\text{max} = 120\,\text{veh}/\text{km}/\text{lane}$ and above. Clearly, this is impossible with functions $Q(\rho)$ that are concave down. However, when admitting inflection points, it is possible. The resulting traffic models with non-convex flux functions will generate interesting new phenomena (cf.~\cite{LeVeque2002}), most prominently mixed waves that are composed of rarefactions and shocks.

An important aspect to address is the fundamental shortcoming of the ARZ model, outlined in point 2) in \S\ref{sec:results_model_accuracy}, namely that the different curves $Q_w(\rho)$ do not possess a uniform density at which the velocity vanishes. Due to the form \eqref{eq:arz_generalized_Q} of the curve family, this is impossible within the ARZ framework. A way out is provided by the framework of generic second order models \cite{LebacqueMammarHajSalem2007} that lead to generalized ARZ models, as presented in an upcoming paper \cite{FanHertySeibold2013}. Another aspect of interest is the study of second-order models that possess a relaxation term in the momentum equation (cf.~\cite{AwRascle2000}). As pointed out in \cite{Greenberg2004, FlynnKasimovNaveRosalesSeibold2009, SeiboldFlynnKasimovRosales2013}, the presence of a relaxation term can reproduce a variety of wave features that are observed in real traffic flow.

\vspace{1.5em}
\section*{Acknowledgments}
The authors would like to thank Michael Herty, Andrew Moylan, Benedetto Piccoli, Rodolfo Ruben Rosales, and Daniel B. Work for helpful comments and discussions. B. Seibold would like to acknowledge the support by the National Science Foundation though grant DMS--1115269, and partial support through grants DMS--1318641 and DMS--1318709. This research was supported in part by the National Science Foundation through major research instrumentation grant number CNS--09--58854.

\vspace{1.5em}
\bibliographystyle{plain}
\bibliography{references_complete}

\end{document}